\documentclass[12pt]{article}

\usepackage[utf8]{inputenc}
\usepackage{amsmath,amsfonts,amsthm}
\usepackage{graphicx,psfrag,epsf,epstopdf}
\usepackage{enumerate}
\usepackage{bbm}
\usepackage{url,hyperref}

\usepackage{booktabs}
\usepackage[table,xcdraw]{xcolor}
\usepackage[font=small]{caption}
\usepackage{booktabs}
\usepackage{extarrows}
\usepackage{epsf,epsfig,color}
\usepackage{algorithm}
\usepackage{algpseudocode}

\newtheorem{theorem}{Theorem}[section]
\newtheorem{lemma}[theorem]{Lemma}

\newtheorem{prop}{Proposition}

\newcommand{\E}{\mathbb {E}}

\newcommand{\M}{\mathbb {M}}
\newcommand{\0}{{\bf 0}}
\newcommand{\tod}{\stackrel{{\cal D}}{\longrightarrow}}
\newcommand{\toP}{\stackrel{{\cal P}}{\longrightarrow}}
\newcommand{\eqd}{\stackrel{{\cal D}}{=}}
\newcommand{\Var}{{\rm Var}}

\def\PP{\mathbb{P}}
\def\R{\mathbb{R}}

\def\N{\mathbb{N}}
\def\M{\mathbb{M}}
\def\X{{\cal X }}
\def\P{{\cal P }}
\def\be{\begin{equation}}
\def\ee{\end{equation}}

\def\ka{{\kappa}}
\def\qed{\hfill\hbox{${\vcenter{\vbox{ \hrule height 0.4pt\hbox{\vrule width 0.4pt height 6pt \kern5pt\vrule width 0.4pt}\hrule height 0.4pt}}}$}}
\def\spacingset#1{\renewcommand{\baselinestretch} {#1}\small\normalsize} \spacingset{1}

\renewcommand{\baselinestretch}{1.2}

\providecommand{\keywords}[1]{\small\textbf{\textit{Keywords---}} #1}

\title{\bf Localization processes for functional data analysis}
\author{Antonio Elías$^{1,2}$, Raúl Jiménez$^{1,2}$ and J. E. Yukich$^{1, 3}$\footnote{Corresponding author. Email:  \url{joseph.yukich@lehigh.edu}}  \\
	\small $^{1}$ Department of Statistics, Universidad Carlos III de Madrid \\
	\small $^{2}$ UC3M-Santander Big Data Institute, Universidad Carlos III de Madrid \\
	\small $^{3}$ Department of Mathematics, Lehigh University \\
}
\date{\today}

\begin{document}

\maketitle

\begin{abstract}
We propose an alternative to $k$-nearest neighbors for functional data whereby the approximating neighboring curves are piecewise functions built from a functional sample.
Using a locally defined distance function that satisfies stabilization criteria, we 
establish pointwise and global approximation results in function spaces when the number of data curves is large enough. 
We exploit this feature to develop the asymptotic theory when a finite number of curves is observed at time-points given by an i.i.d. sample whose cardinality increases up to infinity.
We use these results to investigate the problem of estimating unobserved segments of a partially observed functional data sample as well as to study the problem of functional classification and outlier detection.  For such problems our methods are competitive with and sometimes superior to benchmark predictions in the field.
\end{abstract} 

\hspace{10pt}


\newpage


\section{Introduction}

The $k$-nearest neighbors ($k$NN) method has been identified by IEEE as one of the top algorithms for solving multivariate statistical problems on large datasets \cite{top10}.
It is particularly useful for classification and regression, where the method is based on the idea that similar patterns must  belong to the same class and near explanatory variables will have similar response variables.
Among other applications of $k$NN in the multivariate setting, we also include clustering \cite{Brito}, outlier detection \cite{outliers} and time series forecasting \cite{knnTS}.
Beyond its effectiveness, the popularity of the method is due in part to its conceptual ease  and  implementation.
This has sparked the interest of   researchers, who over many years have developed not only applications but also the mathematical theory, making the method an essential tool in nonparametric multivariate statistics.
A critical review of the seminal literature on the asymptotic theory related to the application of the method to classification, regression and density estimation is provided by Chapter 6 of \cite{GKKW02}.

\bigskip
In the context of functional data analysis (FDA), the $k$NN method has also been explored.
For example, \cite{ZJS10} addresses the problem of forecasting final prices of auctions via functional $k$NN and \cite{HRS17} constructs classifiers for functional data also based on $k$NN.
However, the asymptotic theory of these methods has remained undeveloped until now.
Asymptotic results of methods based on functional $k$NN are mainly related to regression estimation when the response variable has finite dimension \cite{BCG10, KV13}.
Some of these results have been extended to other operations on  the response variable \cite{KLRV17}, such as conditional distribution and conditional hazard function.
And, exceptionally,  \cite{Lian11} proves the consistency of some regression estimates based on  $k$NN when both dependent and independent variables are functions.
Thus, in the functional data setting, the mathematical theory of the $k$NN rule is relatively
unexplored.

\bigskip
One of the purposes of this paper is to fill this lacuna and to provide the asymptotic theory for methods  inspired by $k$NN. 
The proofs of the main results rely on the theory of stabilizing functionals.
This theory has been mainly developed for establishing limit theorems for statistics arising in stochastic geometry \cite{Schreiber10}.
At its core, this theory is applicable to statistics which are expressible as sums of score functions which depend on local data in a well-defined way.
Statistics involving multivariate $k$NN are prime examples of locally defined score functions.
In this context we introduce a $k$th localization process which well approximates a target
process in both a pointwise and $L^1$ sense.
We exploit the fact that the $k$th localization process is locally defined to rigorously develop its  first and second order limit theory.

\bigskip
The second purpose of this paper is to review some problems of the FDA literature from a perspective enriched by the new asymptotic results.
In particular, we consider the  problem of estimating unobserved values of a partially observed functional data sample, 
a question prominently addressed in the literature \cite{Kraus15, YMW12}.
As has been already reported \cite{ZJS10}, the  $k$NN method is a natural approach for addressing this problem.
We show that the method may provide estimates that are superior to benchmark predictions in the field and we provide regularity conditions for guaranteeing their consistency.
We also consider the problems of classification and outlier detection.
Specifically, we introduce a probabilistic functional classifier inspired by the $k$NN rule.
The classification method is based on the asymptotic normality of an empirical distance which  only considers a finite number of sample curves but takes advantage of the 
fact that the data live in a function space, giving rise to empirical 
methods in such spaces.
As far as we know, this type of asymptotic result is new and is particular to the functional setting.
The problem of outlier detection is straightforwardly tackled by considering one-class classification.
We carry out a comparative study with other widely used methods in FDA \cite{AR14,  LCL12, SG11} which  shows that our approach performs well across different test datasets.
In addition, with the new classifier we may predict classification probabilities rather than only outputting the most likely class.

\bigskip


\subsection{Definitions and terminology}

Functional data are typically viewed as independent realizations of a stochastic process  with smooth trajectories observed on a compact interval \cite{YMW12}. Consequently,
 we consider a stochastic process $X=\{X(t):t\in[a,b]\}$ with continuous sample paths. Following standard practice we set $[a,b] = [0,1]$.
Let $X_1,\dots,X_n$ be independent copies of the process $X= X(t), t \in [0,1]$. The $X_i, i \geq 1,$ take values in $C[0,1]$, the space of continuous functions on $[0,1]$.
The classical way of defining a distance between sample paths $X_i$ and $X_j$ is to use the $p$-norm, $p\geq 1$, also called the Minkowski distance,
$$
D(X_i,X_j) = \Big(\int_0^1 |X_i(t) - X_j(t)|^p \Big)^{1/p} dt.
$$
Other distances, such as the Hausdorff  distance, may be defined in terms of distances between two nonaligned points of the curves  $X_i$ and $X_j$. By two nonaligned points, we mean  $(t, X_i(t))$ and $(s, X_j(s))$ with $t\neq s$.
Such distances are not considered in the present work.
In any case, given a distance $D(\cdot,\cdot)$ between two functions, the global nearest neighbor to $X_i$  is defined by
$$
X_i^{(1)}= \arg \min_{\{X_j:j\neq i\}} D(X_i,X_j).
$$
Iterating, for $k \in \{2,...,n - 1\}$, 
 the global $k$NN to $X_i$ is defined as  the nearest neighbor in the subsample  $\{X_1,\dots,X_n\} \setminus \big\{X_i^{(1)}, \dots, X_i^{(k-1)}\big\}$.

\bigskip
In this paper we  consider {\em local} nearest neighbors to a given curve 
which without loss of generality is taken to be $X_1$. 
We begin by considering
the nearest function to $X_1$ in the pointwise sense. This gives rise to the stochastic process
$$
\hat{X}_1^{(1)}(t) = 
\hat{X}_{n,1}^{(1)}(t)
 = \arg \min_{{\{X_j(t):j\neq 1\}}} |X_1(t)-X_j(t)|, \ \ t\in [0,1],
$$
which we call the {\em first localization process.}
This process consists of a union of
sample curve segments which are the nearest sample observations to $X_1$.
We define the $k$-nearest sample piecewise function to $X_1$ by iterating, in way similar to how we defined the global $k$NN:
First, for $t\in [0,1]$, let $G_1^{0}(t) = \{X_j(t):j\neq 1\}$. Then, for $1 \leq k\leq n -1$, define
$G_1^k(t) = G_1^{k-1}(t)\setminus \{\hat{X}_1^{(k)}(t)\}$. Thus, we define the  {\em $k$th localization process} by
\begin{equation}
\label{lprocess}
\hat{X}_1^{(k)}(t)  = \hat{X}_{n,1}^{(k)}(t) = 
 \arg \min_{x(t) \in G_1^{k-1}(t)} |X_1(t)- x(t)|, \ \ t \in [0,1].
\end{equation}
This curve is the central object of our studies.  We will show that it well approximates
$X_1$ in a pointwise and global sense.

The localization distances between $X_1(t)$ and $\hat{X}_i^{(k)}(t),  t \in [0,1]$, give rise to the {\em $k$th localization width process}
$$
L^{(k)}_n(t)= L^{(k)} (X_1(t),  \{X_j(t) \}_{j = 1}^n )  = \big|X_1(t) - \hat{X}_1^{(k)}(t)\big|,\ t \in [0,1].
$$
The R package \texttt{localFDA} at \url{https://github.com/aefdz/localFDA} provides the programs for computing the localization and localization width processes.  Under mild conditions on the
marginal density of $X(t)$, we shall show that the $L^1$ norm of $L^{(k)}_n(\cdot)$
 is $O\big( k/2n \big)$.
Hereinafter, we assume that the marginal probability density of $X(t)$, denoted by $\ka_t$, exists for almost all   $t \in [0,1]$. 
Denote by $S(\kappa_t)$  the possibly unbounded support of 
$\ka_t$.  
 
 We will assess the closeness of the $k$th localization process  $\hat{X}_1^{(k)}(t),
 t \in [0,1],$ to the data curve $X_1(t), t \in [0,1],$ by studying the   \emph{re-scaled localization width process} 
\be
\label{width}
W_n^{(k)}(t)= 
W^{(k)}\Big(X_{1}(t),  \{X_j(t) \}_{j = 1}^n\Big) = \frac{2n} {k} L^{(k)}_n(t), \ \ t \in [0,1].
\ee
The re-scaled localization distance $W_n^{(k)}$ is invariant under any affine transformation of the data, i.e.,  transformations of the data by  functions of the type  $T(x) = ax + b$  leave $W_n^{(k)}$  unchanged.
Both the results and methods discussed in this paper are invariant under affine transformations.

\subsection{Outline of this work}
This paper is organized as follows.
Section 2 presents three types of asymptotic results.
\begin{itemize}
\item[(a)] {\em Pointwise convergence.}  
We show  mean and distributional convergence of $W_n^{(k)}(t)$, at a fixed $t \in [0,1]$, when $n \to \infty$. 
\item[(b)]  {\em Process convergence.}  Under regularity conditions on the density of the data, the  $L^1$ norm of  the average difference between ${W}_n^{(k)}(t),  t \in [0,1],$ and a limit localization process is $o(1)$.   
This yields that the expected $L^1$ norm of 
$\big|X_1(t) - \hat{X}_1^{(k)}(t)\big|$ is $O(k/2n)$.


\item[(c)] {\em Asymptotic normality}. For a fixed number of curves $n$,
fixed $k$, and for ${\cal T}_m$ a set of $m$ i.i.d.  locations, the sum $\sum_{t\in {\cal T}_m}L^{(k)}_n(t) $ follows a Gaussian distribution when  $m$ increases up to infinity.  
\end{itemize}
 Limit results (a) and (b) are used in Section 3, where we consider estimation of missing values of partially observed data via $k$NN.
 Limit result (c) is used in Section 4, which provides a new method for classification and outlier detection.
A general discussion of both theoretical and practical results is given in Section 5 whereas  Section 6 provides the proofs of our main results of Section 2.

\section{Main results} \label{Sec2}

\subsection{Asymptotics for localization processes with large data sizes }

We first assess the pointwise behavior of the re-scaled localization width process.  Let
$|A|$ denote the Lebesgue measure of the set $A$.

\begin{theorem}  
\label{mainthm1} For all integers $k \in \N$ and  almost all $t \in  [0,1]$, we have
\be \label{WLLN}
\lim_{n \to \infty} \E W_n^{(k)}(t)  = |S(\kappa_t)|
\in (0, \infty]
\ee
and
\begin{align} \label{Var}
\lim_{n \to \infty}  \Var [  W_n^{(k)}(t)] = \Big(1 + \frac{1} {k}\Big)   \int_{S(\kappa_t)} \frac{  1 } {\ka_t(y) } dy - | S(\kappa_t)|^2 \in
 \left[ \frac{|S(\kappa_t)|^2}{k}, \ \infty\right].
\end{align}
In addition, as $n \to \infty$,
\be
\label{WLLN2}
 W_n^{(k)}(t) \tod  W_{\infty}^{(k)}(t) = 
  \frac{2}  {k}  \frac{1} { \ka_t(X_1(t))} \Gamma(k,2),
  \ee
  where $\Gamma(k,2)$ is a Gamma random variable with shape parameter $k$ and scale parameter
  $2$. 
\end{theorem}

\bigskip

The Jensen inequality applied to the mapping $x \to x^{-1}$ shows that
$\frac{1} {| S(\kappa_t)|}   \int_{S(\kappa_t)} \frac{  1 } {\ka_t(y) } dy \geq | S(\kappa_t)|$.  Thus the asympotic variance in  \eqref{Var} is greater than or equal to  $\frac{1}{k}| S(\kappa_t)|^2$ and attains equality when $\ka_t$ is the uniform density on $S(\kappa_t)$. 

\bigskip
We next assess the global behavior of the localization process in the $L^1$ norm on $C[0,1].$
We find conditions under which  the $L^1$ norm of the expected  difference between the localization width process $W_n^{(k)}(t)$ and $W_{\infty}^{(k)}(t)$ defined at \eqref{WLLN2}
converges to zero.  
Thus, the target function $X_1$ is {\em globally} well approximated by the localization process.
Specifically, the asymptotic expected
$L^1$ error in locating a typical curve by its $k$th localization process is $O(k \int_0^1 |S(\kappa_t)| dt/2n) $, which depends on $k$ and on the Lebesque measure of the support of the underlying distribution of the data.  This happens provided that the data is {\em regular from below}, i.e., for almost all $t \in [0,1]$ we have $|S(\kappa_t)| < \infty$ and 
there exists $\ka_{\rm{min}}>0$ and an interval $S_\delta \subset S(\kappa_t)$, with $|S_\delta | = \delta>0$, such that
\be \label{reg}
\inf_{x \in S_\delta} \ka_t(x) \geq \ka_{\rm{min}}.
\ee
Examples of data which are regular from below include  harmonic signals of the form $X(t) = A \sin(2\pi t) + B \cos(2\pi t)$, where the coefﬁcients $A$ and $B$ are independent random variables having  a uniform distribution on some  compact interval. 
These processes have been used in simulation studies by several authors; e.g. by \cite{HS2010} and \cite{SG11}, among others.
More generally,  finite Fourier sums with independent random coefficients having a probability density defined on a compact interval are also examples of processes satisfying \eqref{reg}.
Due to either physical or biological restrictions, or to limitations of supply,  functional data are often  bounded. This includes, for example,  mortality, fertility and migration rates \cite{Hyndman2007, Hyndman2008}; curves of surface air temperatures and precipitation \cite{Genton18}; electricity market data \cite{Liebl19} and functional data from  medical studies \cite{Kraus15, YMW12}.
For such data, we may assume the data is regular from below.  The 
next two process level results apply to such data curves.

\begin{theorem}  \label{global} 
Assume  that the data is regular from below as at \eqref{reg}. Then
\be
\lim_{n \to \infty}   \int_0^1  \E \left| {W}_n^{(k)}(t) - {W}_{\infty}^{(k)}(t) \right| dt= 0.
\ee
Consequently $\lim_{n \to \infty}   \int_0^1\E W_n^{(k)}(t) dt =  \int_0^1 |S(\kappa_t)| dt
$ and the average $L^1$ error satisfies
\be \label{star}
\E \int_0^1 \big|X_1(t) - \hat{X}_1^{(k)}(t)\big| dt
 \leq \frac{k}{2n} \left( o(1) + \int_0^1 |S(\kappa_t)| dt\right) = O\left( \frac{k}{2n}\right).
\ee
\end{theorem}

 
\bigskip

Under further conditions we may extend the convergence \eqref{star} to curves
other than $\hat{X}_1^{(k)}$.   This will be spelled out in 
more detail in Proposition \ref{propo}  in Section 3. To prepare for this, we  consider the case when $k= k(n)$ increases with $n$.
This result will be used in Section 3 to aid in reconstructing a curve when data is missing.

\begin{theorem} \label{mainthm2}
Assume  that the data is regular from below as at \eqref{reg}.
Let $\kappa_t$ be $\alpha$-H\"older continuous, i.e., there is $\alpha \in (0,1]$ such that
$$
|\kappa_t(x) - \kappa_t(y) | \leq C |x - y|^{\alpha}, \ \ x,y \in S(\kappa_t).
$$
 Let $k= k(n)$ satisfy
 $\lim_{n \to \infty} \frac{ k^{1 + \alpha}}{n^{\alpha}} = 0$.  Then
\be \label{WLLNk}
\lim_{n \to \infty}  \E W_n^{(k)}(t) =  |S(\kappa_t)| \in (0, \infty]
\ee
and 
\begin{align} \label{Vark}
 \lim_{n \to \infty}  \Var [ W_n^{(k)}(t)] =  \int_{S(\kappa_t)} \frac{  1 } {\ka_t(y) } dy -  |S(\kappa_t)|^2 \in [0, \infty].
\end{align}
Additionally, 
\be \label{WLLNkLone}
\lim_{n \to \infty}  \int_0^1\E W_n^{(k)}(t) dt = \int_0^1 |S(\kappa_t)| dt.
\ee
\end{theorem}

\bigskip
The right-hand side of \eqref{Vark}  vanishes if $\kappa_t$ is uniform on its support.
In this  case  $W_n^{(k)}(t)$ converges to $ |S(\kappa_t)|$ 
in probability as $n \to \infty$.

\subsection{Stochastic behavior of empirical localization distances}
One often observes functional data on a discrete time set $\{t_1,   \dots , t_m\}$.
In such cases, the average distance between $X_1$ and its $k$th localization process \eqref{lprocess} is a global measure of nearness.
This distance is given by the average width
\be \label{ave}
\frac{1}{m} \sum_{r=1}^m L^{(k)} (X_1(t_r),  \{X_j(t_r) \}_{j = 1}^n ).
\ee

\bigskip
Here we focus on the distributional behavior of \eqref{ave}  when $\{t_1,   \dots , t_m\}$ is the realization of i.i.d. uniform random variables $T_1,...,T_m$ on $[0,1]$.
This gives rise to an empirical localization width process and goes as follows.
  We fix $n$, the number of data curves.  
We evaluate the localization distance with respect to $X_1$ at each $T_r, 1 \leq r \leq m$.
This generates the so-called {\em empirical localization distance}  between $X_1$ and its $k$th localization process namely
\be \label{empirical}
\frac{1} {m} \sum_{r = 1}^m   L^{(k)} (X_1(T_r),  \{X_j(T_r) \}_{j = 1}^n ) = 
\frac{1} {m} \sum_{r = 1}^m   L^{(k)}_n (T_r).
\ee
The empirical localization distance is simply the localization width $L^{(k)}_n(\cdot)$ averaged over  the sample  $\{T_r\}_{r = 1}^m$.

\bigskip
The localization widths  $ L^{(k)}_n (T_r)$, $1 \leq r \leq m$,   exhibit dependence in general. However, if $L^{(k)}_n (t')$  depends only on the values of $L^{(k)}_n (t)$  at 
preceding data points $t \in \{T_r\}_{r = 1}^m$
within distance $\frac{M}{\sqrt{m}}$ of $t'$,  where $M$ is a fixed positive constant, then   the asymptotic normality
 as $m \to \infty$  of the empirical localization distance  follows from 
 $M$-dependence as follows.  Such a dependency assumption holds if the data has a 
 Markovian structure, with $X(t)$  depending only on the immediate past
 $X(t^{-})$.

\bigskip

\begin{theorem}  \label{CLT01} Fix $n$, the number of data functions.
 Assume that the localization distances $ L^{(k)}_n (t), t \in \{T_r\}_{r = 1}^m$, depend only on the values of $ L^{(k)}_n (\cdot)$ at preceding data points $t \in \{T_r\}_{r = 1}^m$
within distance $\frac{M}{\sqrt{m}}$ of $t$, where $M$ is a fixed positive constant.
Then as $m \to \infty$ we have
\be
\label{eqtheo2.4}
 \frac{  \sum_{r = 1}^{m}  ( L^{(k)}_n (T_r) - \E L^{(k)}_n (T_r) )  }
{  \sqrt{ \Var  [ \sum_{r = 1}^{m}  L^{(k)}_n  (T_r)] }   }  \tod N(0, 1).
\ee
\end{theorem}

\bigskip
The expected value and the variance in (\ref{eqtheo2.4}) may be estimated by sample means and sample variances of empirical localization distances. 
To achieve this, we must observe the empirical localization distance to each sample curve, not only to $X_1$.
We compute the empirical localization distance to $X_i$  by replacing  $X_1$  with  $X_i$ in \eqref{empirical}.
Let $L_i^{(k)}$ be the corresponding statistics. 
The sample mean and sample variance are
$$
\bar{L}^{(k)}= \frac{1}{n}\sum_{i=1}^n L_i^{(k)} \ \ \mbox{and} \ \  S_L^2 = \frac{1}{n-1}\sum_{i=1}^n (L_i^{(k)} - \bar{L}^{k})^2.
$$
When $n$ is large, $T_i^{(k)} =(L_i^{(k)}-\bar{L}^{(k)})/S_L$  is approximately distributed as the centered and normalized empirical localization distance  on  the left-hand side of (\ref{eqtheo2.4}) .
Thus, Theorem 2.4 suggests that the statistics $T_i^{(k)}$ could be used for testing whether \emph{$X_i$ is properly localized by the data} in accordance with an underlying Gaussian distribution.
We explore this idea for classification and outlier detection in Section 4.

\section{Reconstruction of partially observed data via $k$NN}

One might hope that two  curves which are near on a set $S \subset [0,1]$ and which are copies of the same process should remain near on $[0,1]\setminus S$. In particular, if $S$ is blindly chosen and with large Lebesgue measure, one could hope to achieve this proximity without taking into account prior morphological  information, such as shape and complexity,  of the sample curves. 
Here we show that this turns out to be the case, subject to mild assumptions on the data.  
This is achieved by making  use of $k$-nearest neighbor methods for reconstructing partially observed data.

\bigskip
As is customary in the literature, we model partially observed functional data   
by considering a random mechanism  $Q$ that  generates compact subsets of $[0,1]$.
These sets correspond to ranges where sample paths are observed. 
Formally, $O_1, \dots, O_n$ are independent random closed sets from $Q$ such that  $X_i$  is observed on $O_i$ and is missed on  $M_i = [0,1]\setminus O_i$.
We also will assume data are Missing-Completely-At-Random, i.e., the sets $\{O_i\}$ are independent of the sample paths \cite{KL20}.  
Without loss of generality, suppose also that there is no time which is almost surely censured.
That is to say we assume  $\PP( O_i \ \mbox{contains} \ s  ) > 0$  for almost all $s\in [0,1]$.

\bigskip
To  simplify the notation, consider first the case in which just one sample path is partially observed, say $X_1$.
Therefore we assume for now that $X_2,\dots,X_n$ are fully observed on $[0,1]$.
Instead of the Minkowski distances  to $X_1$
taken on the complete observation range $[0,1]$, we now consider such distances  restricted to $O_1$, namely
\be
\label{Dp}
D_p(X_j) = \bigg(\int_{O_1} |X_j(t) - X_1(t)|^p \bigg)^{1/p} dt.
\ee
For $1 \leq j \leq n -1$, denote by $X^{(j)}$ the $j$NN to $X_1$ with respect to this distance.
To estimate $X_1$ on $M_1$, we adopt the $k$NN methodology  and consider  convex combinations of the form $\sum_{j=1}^r w_j\cdot  X^{(j)}$.
The choice of $r$ and suitable weights $\{w_j\}_{j=1}^r$ will be discussed later.
We start by providing conditions for the consistency of this type of estimator.
For this, it is enough to discuss conditions for the consistency of  $X^{(j)}$ as an estimator of $X_1$.

\bigskip
Choose an arbitrary $l\neq 1$ and consider the random interval  on which $X_l(t) $ is closer to $X_1(t)$ 
than is the $k$th localization process $\hat{X}_1^{(k)}(t)$.
That is to say we consider the random interval 
\be \label{set}
I^{(k)}(X_l) = \Big\{t\in [0,1]: \big|X_l(t) -X_1(t)\big| \leq \big|\hat{X}_1^{(k)}(t) -X_1(t)\big|\Big\}.
\ee
Note  $\PP( I^{(k)}(X_l) \  \mbox{contains} \ s  )  = k/(n-1)$ for all $s \in [0,1]$.
Since $X^{(j)}$ is selected by its proximity to $X_1$ on $O_1$,
and  since $O_1$ is independent of $X_1$ and $\hat{X}_1^{(k)}$,  one expects, as $k$ increases up to $n -1$, that 
$\PP( I^{(k)}\big(X^{(j)}\big) \ \mbox{contains} \ s )$
increases up to $1$ faster than
$\PP( I^{(k)}(X_l) \  \mbox{contains} \ s  )$
for any fixed $j$ and $s \in [0,1].$
More formally, we will consider the following assumption:
\be 
\label{ass22}
\lim_{n\rightarrow \infty} \PP\big( I^{(k)}\big(X^{(j)}\big) \ \mbox{contains} \ s  \big) = 1   \ \ \mbox{for some} \ \ k=o(\sqrt{n}),  \ \ s \in [0,1] .
\ee

\begin{figure}[!t]
	\includegraphics{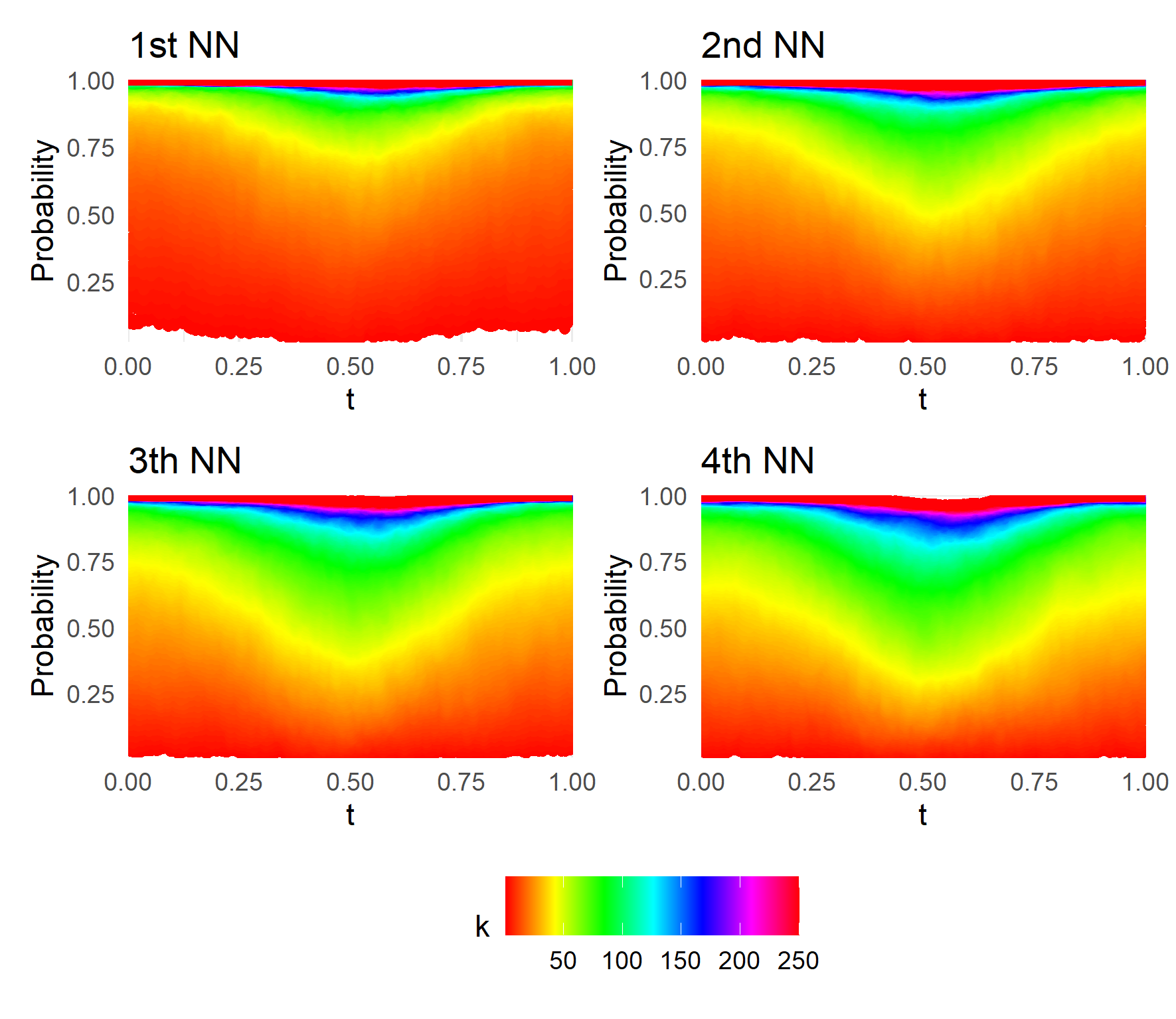}
	\caption{Estimated values of $\PP\big( I^{(k)}\big(X^{(j)}\big)  \ \mbox{contains} \ t \big)$, $0\leq t \leq 1$, $1\leq j\leq 4$, $1 \leq k\leq 250$, $n=2500$.
	The estimation is based on $1000$ independent replicates of $(X_1,O_1)$  when  $O_1$ is obtaining by removing randomly one interval of the partition of $[0,1]$ induced by two independent Uniform(0,1) variables. $X_1$ is a linear combination of sines and cosines with independent Gaussian coefficients.}
\label{fig:probabilities_fourier}
\end{figure}

Given the features of many functional data used in practice, condition (\ref{ass22}) does not appear too unusual.  
In many cases, the functions are smoothed data 
by Fourier analysis. 
This is the reason why simulation studies often consider Fourier sums with random coefficients for generating test data.
In this context we note that Fourier sums are close to a target curve whenever the respective coefficients are close.  Moreover, if the Fourier sums are close to a target on an observable window in $[0,1]$, then they are close everywhere in $[0,1]$, since the coefficients do not depend on $t$. 
In such cases, if $j$ is fixed 
the $j$NN is on average at distance $O(1/n)$.
On the other hand, if  $k=o(\sqrt{n})$, the $k$th localization process is at a 
 distance $O( k/n) = o(1/\sqrt{n})$, verifying (\ref{ass22}).
As an illustration, Figure~\ref{fig:probabilities_fourier} shows empirical estimators of $\PP\big( I^{(k)}\big(X^{(j)}\big)    \ \mbox{contains} \ s  \big)$ based on $1000$ replicates of $(X_1,O_1)$,  when $n=2500$ and $k$ 
ranges over integers
up to 250. $O_1$ is obtaining by removing at random one of the three closed intervals of the subdivision of $[0,1]$ induced by two independent Uniform(0,1) random variables.
$X_i$ is a linear combination of sines and cosines with independent normal coefficients, as those used for generating data in previous studies \cite{KL20,  Kraus15}.

\bigskip
\begin{prop} \label{propo}  Assume the data is regular from below
as at \eqref{reg}. 
Suppose $\kappa_t$ is $\alpha$-H\"older continuous
with $\alpha = 1$
 and that (\ref{ass22}) holds. Then for any fixed $j$ and all $\varepsilon>0$   we have
$$
\lim_{n\rightarrow \infty} \PP\big(  \int_0^1 |X^{(j)}(t) -X_1(t)| dt <\varepsilon\big) =  1.
$$
\end{prop}

\noindent{\em Proof. }  We have
\begin{align}
 \PP\big(  \int_0^1 |X^{(j)}(t) -X_1(t)| dt >\varepsilon\big) & \leq 
 \PP\big(  \int_0^1 |X^{(j)}(t) -X_1(t)| {\bf 1}_{I^{(k)}(X^{(j)})}(t) dt  > \frac{\varepsilon}{2} \big) 
 \nonumber \\
 & \hskip.5cm + \PP\big(  \int_0^1 |X^{(j)}(t) -X_1(t)| {\bf 1}_{[0,1] \setminus I^{(k)}(X^{(j)})} (t) dt  >\frac{\varepsilon}{2} \big) \nonumber \\
 & \leq  \frac{2}{\varepsilon}  \E \int_0^1 |\hat{X}_1^{(k)}(t) -X_1(t) | dt +
  \E \int_0^1 {\bf 1}_{[0,1] \setminus I^{(k)}(X^{(j)})} (t) dt \nonumber \\
 & = \frac{k}{n\varepsilon} \E \int_0^1  W_n^{(k)}(t) dt + \E \int_0^1 {\bf 1}_{[0,1] \setminus I^{(k)}(X^{(j)})} (t) dt .
 \end{align}
Since $\kappa_t$ is $\alpha$-H\"older continuous,  we may apply  Theorem \ref{mainthm2} for $\alpha = 1$ and $k= k(n) = o(\sqrt{n})$.
Thus, as $n \to \infty$,  the right-hand side goes to 0 by Theorem \ref{mainthm2},
the finiteness of $\E \int_0^1  W_n^{(k)}(t) dt$,  and (\ref{ass22}).  \qed

\subsection{The functional $kNN$ method}

As is customary, we suppose there is a proportion 
of curves completely observed.
In this case, we may repeat the above approach for estimating any sample curve partially observed by choosing its  $j$NN from among the curves which are fully observed.
Moreover, as we remarked,  if there is a significantly large proportion of curves with missed values at $t$, then Theorem \ref{CLT} provides asymptotic confidence intervals for average errors when  imputing missing values by localization processes.
In view of our construction, if $k$ is large but finite, these errors may be used for bounding, with significantly high probability, errors resulting when estimating missing values by nearest neighbors.
In other words, if \eqref{set} holds, based on Theorem  \ref{CLT} in the Appendix
which gives rates of normal convergence
 for $W_n^{(k)}(t)$, we may provide approximate confidence intervals for average errors when estimating by nearest neighbors. 
In particular, following the discussion around \eqref{averageerror} in the Appendix, these average errors are $O(k/n)$ in mean, with variance $O(k^2/n^3)$.
This is an additional attraction of the functional $k$NN estimators method which  we now describe.

\bigskip
Let us go back to the simplest case, where the curve $X_1$ is observed on $O_1$ and unobserved on $M_1$, where 
$X_2,\dots,X_n$ are fully observed on  the entire observation range, and where the $k$NN estimator of $X_1(t)$ has the form
$$\hat{X}_{k\mbox{NN}} = \sum_{j=1}^r w_j\cdot  X^{(j)}(t),$$
with $w_j>0$, for $1\leq j \leq k$, and $\sum_{j=1}^r w_j = 1$.
We follow previous literature \cite{HRS17, ZJS10} and consider  Minkowski distances with $p=1$ and $p=2$.
In the forecasting context, both functional data and univariate time series, the weights for these  Minkowski distances have been already suggested \cite{ZJS10, knnTS}.
We use these recommendations and set $w_j = D_p\big(X^{(j)}\big)^{-p}/\sum_{i=1}^r D_p\big(X^{(i)}\big)^{-p}$, $D_p\big(\cdot)$ being the distance to $X_1$ defined in \eqref{Dp}.
The value of $r$ used to define the $k$NN estimator is chosen by minimizing the mean square error between the estimator and the target function $X_1$ on the observation range. 
This is 
$$
r = \arg \min_{r} \int_{O_1} |\hat{X}_{k\mbox{NN}}(t) - X_1(t)|^2 dt.
$$
The Mean Square Errors on $M_1$ (MSE), that is to say $\int_{M_1} |\hat{X}(t) - X_1(t)|^2 dt$, are used for evaluating the estimator performance.

\bigskip
To illustrate the method, we conducted a simulation study based on two real case studies.
The differences between the results obtained by the functional $k$NN method based on the Minkowski distance with $p=1$ and $p=2$ were negligible, being slightly superior for $p=2$. 
For the purposes of succinctly summarizing the data,  we only report results for $p=2$.

\bigskip
\subsection{Yearly curves of Spanish temperatures}
Yearly curves of daily temperatures are  common in FDA \cite{Genton18, sara:2009, ramsay:2005}. 
We consider $2786$  such curves from $73$ weather stations  located in the capital cities of $50$ Spanish regions (provinces).
The data was obtained from  \url{http://www.aemet.es/}, the Meteorological State Agency of Spain (AEMET) website.
The date at which the data was first recorded varies from station to station. 
For example, the Madrid-Retiro station reports records from $1893$ onwards whereas the Barcelona-Airport started in $1925$ and Ceuta from $2003$. 
On the other hand, it is likely that some states failed at some moment to record data.
The point is that there are several incomplete years \cite{galeano:2019}.
With the aim of estimating the missing data, we test several methods with a simulation study based on this data set.
From the 2786 fully observed curves we selected one at random, labeled as $X_1$, at which we censored a random number of consecutive days.
Term $O_1$  represents the uncensored days.
The average number of censured days was 122, a third of the year.
We repeat this procedure 1000 times and estimate the censured data by using the $k$NN method and the following benchmark methods to which we append acronyms so as  to easily refer to them in what follows:

\begin{enumerate}
\item PACE \cite{YMW12}. This is the most cited nonparametric method to impute missing data of sparse longitudinal data. 
The method is based on estimations of the classical eigenfunctions, eigenvalues and scores of truncated Karhunen-Lo\`eve decompositions.
For implementation, we used the code from the package \texttt{fdapace} \cite{PACE:2020}.
\item KRAUS \cite{Kraus15}. This is a functional linear ridge regression model for completing functional data based on principal component analysis. 
KRAUS estimates scores of a partially observed functions by estimating their best predictions as linear functionals of the observed part of the trajectory.
Then KRAUS uses a functional completion procedure that recovers the missing piece  by using the observed part of the curve.
The code was obtained from the author's website (\url{https://is.muni.cz/www/david.kraus/web_files/papers/partial_fda_code.zip}).
\item KL20  \cite{KL20}. This  reconstruction method belongs to a new class of functional operators which includes the classical regression operators as a special case. The code was obtained from the author's repository  (\url{https://github.com/lidom/ReconstPoFD}).
\end{enumerate}

\begin{figure}[t]
	\includegraphics{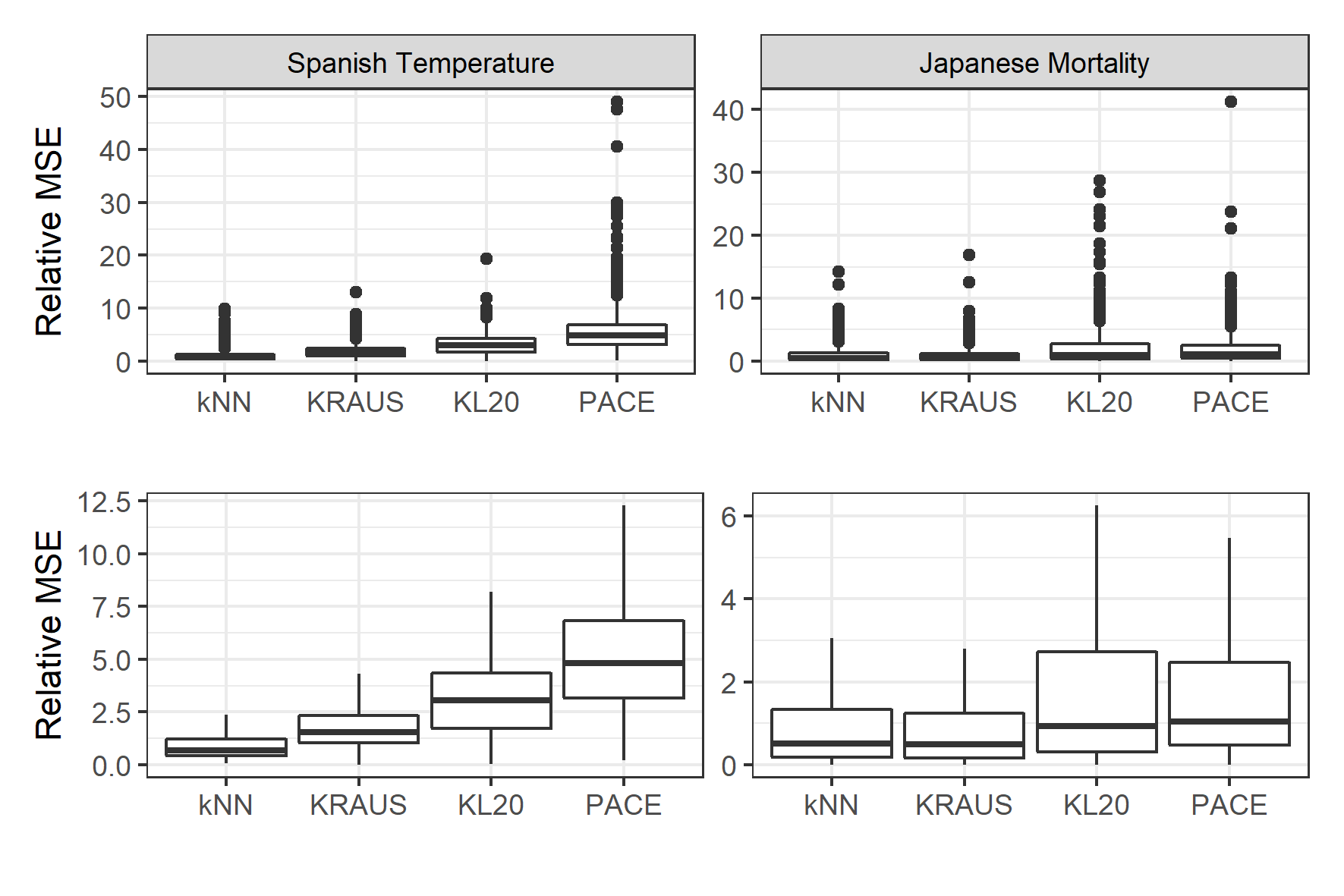}		
	\caption{Top panels: Boxplots of Relative MSE from 1000 reconstruction exercises based on yearly curves of Spanish daily temperatures and Japanese mortality rates. Bottom panels: Blown up images around the median of the above boxplots by excluding atypical values.}
	\label{fig:imputing_mse}
\end{figure}

\begin{table}[]

	\centering
{\small
	\begin{tabular}{@{}lcccc@{}}
		\toprule
		& \multicolumn{1}{l}{kNN}                                   & \multicolumn{1}{l}{KRAUS}                                 & \multicolumn{1}{l}{KL20}                                  & \multicolumn{1}{l}{PACE}                                  \\ \midrule
		Spanish Temperature & \begin{tabular}[c]{@{}c@{}}\textbf{0.1351}\\ \textbf{(0.1134)}\end{tabular} & \begin{tabular}[c]{@{}c@{}}7.4478\\ (1.7490)\end{tabular} & \begin{tabular}[c]{@{}c@{}}3.4935\\ (10.487)\end{tabular} & \begin{tabular}[c]{@{}c@{}}4.0523\\ (0.3510)\end{tabular} \\
		Japanese Mortality    & \begin{tabular}[c]{@{}c@{}}\textbf{0.0197}\\ \textbf{(0.0164)}\end{tabular} & \begin{tabular}[c]{@{}c@{}}0.0980\\ (0.0554)\end{tabular} & \begin{tabular}[c]{@{}c@{}}2.5640\\ (1.1701)\end{tabular} & \begin{tabular}[c]{@{}c@{}}6.6073\\ (3.4661)\end{tabular} \\
		\bottomrule
	\end{tabular}
}
	\caption{Mean running time in seconds observed from 1000 reconstruction exercises based on yearly curves of Spanish daily temperatures and Japanes age-specific mortality rates. Standard deviations are between parentheses.}
	\label{tab:running_times}
\end{table}

\bigskip
By far the best method  was $k$NN. 
For ease in interpreting the results, we report \emph{Relative} MSE, namely 
the  MSE divided by the MSE average when applying the $k$NN method.
This shows that the MSE associated with $k$NN is roughly one half the MSE for KRAUS, a third of the MSE for $KL20$ and a fifth of the MSE for PACE.
This may be observed from the boxplots of Relative MSE on the left side of Figure~\ref{fig:imputing_mse}).
On the bottom of this figure, we zoom in on these boxplots around the median by excluding  the atypical values.
In addition, although all the methods are computationally efficient, $k$NN resulted in being the  most efficient (see Table~\ref{tab:running_times}).

\begin{figure}[!t]
	\includegraphics{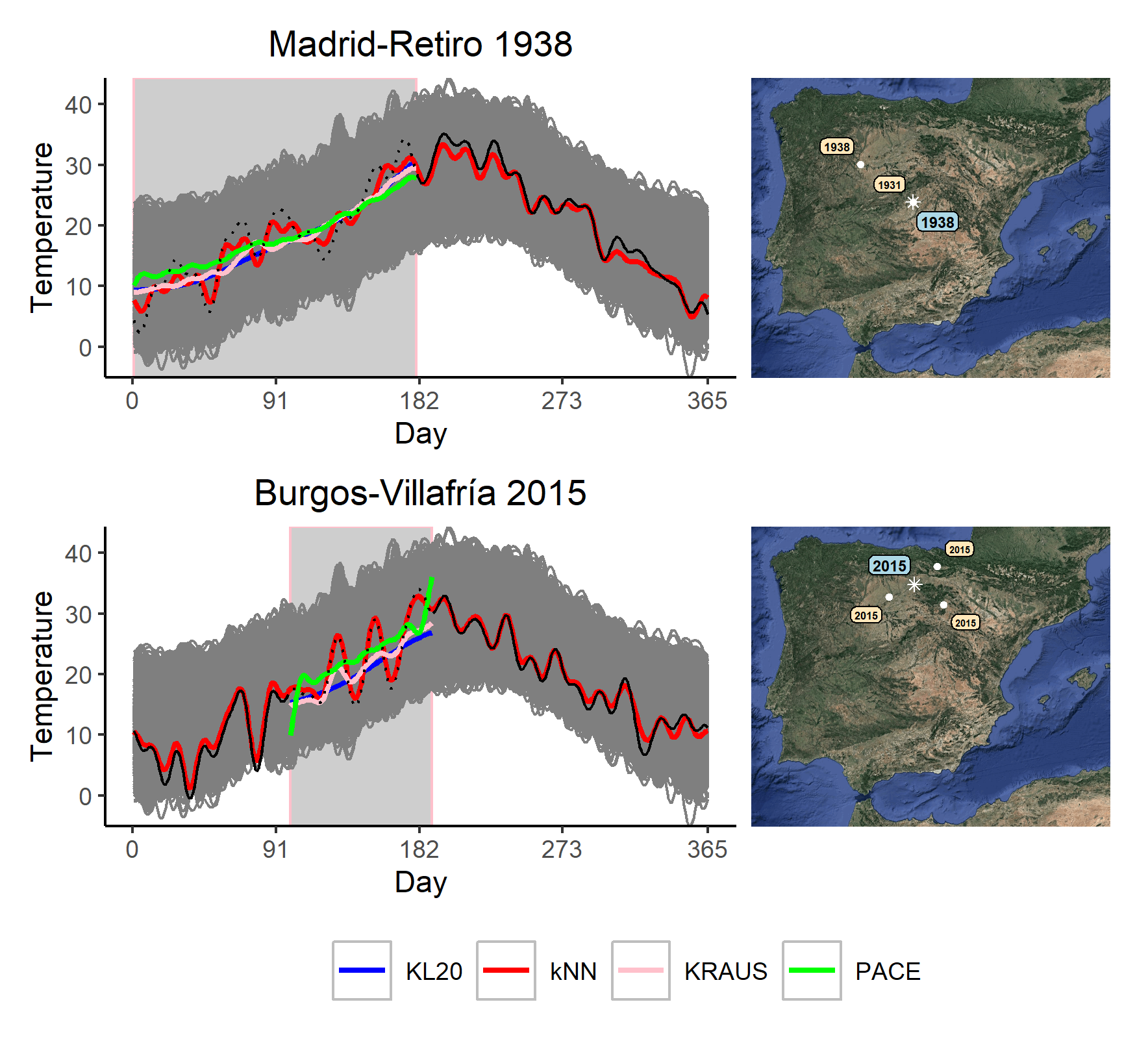}
	\caption{Two illustrations of performance. The randomly observed part of the reconstructed curve is plotted as a black solid line whereas the censored part is dotted. We show both spatial and temporal location of the curves used for reconstruction by $k$NN. For reconstructing Madrid-Retiro 1938, the method chose $k=2$, with Zamora 1938 being the 1st NN and Madrid-Retiro 1931 being the 2nd NN. For reconstructing Burgos-Villafr\'{i}a 2015,  the method chose $k=3$, with Palencia-Autilla Pino 2015 being the 1st NN, Foronda-Txokiza 2015 the 2nd NN, and Soria 2015 the 3rd NN.}
	\label{fig:aemet}
\end{figure}

\bigskip
Figure~\ref{fig:aemet} (left panels) illustrates the typical performance of each method. 
Although all the methods estimate correctly the mean temperature over the daily range, their estimated curve may be  somewhat flattened, without the typical oscillations of Spanish daily temperatures.
Only the $k$NN method provide estimators that may catch both the values of the curve and its shape. 
An additional attraction  of $k$NN is its easy interpretation.
The right panels of Figure~\ref{fig:aemet} depict the $k$NN of the curve under reconstruction, showing that the curves used for reconstruction  come from stations sharing similar weather in a roughly similar time period.

\subsection{Japanese age-specific mortality rates}
The Human Mortality Database (\url{https://www.mortality.org}) provides detailed mortality and population data of $41$ countries or areas.  
For some countries, they also offer micro information by subdivision of the territory,  providing data  rich in spatio-temporal information.
A FDA approach to analyze mortality data is to consider age-specific mortality rates as sample functions  \cite{gao2019, hanlin2017}.
In particular, the Japanese mortality dataset is available for its $47$ prefectures in many years.
However, the curves of some prefectures are incomplete during some years.
In total, we obtained  2007 complete curves of Japanese age-specific mortality rates with data between 1975 and 2016.
We used these curves for comparing the reconstruction methods under consideration by repeating the simulation setup described in the previous subsection.
All the methods perform well on these data.
Typically they do not have the strong oscillations exhibited by  the Spanish temperatures (see the supplementary material for some illustrations).
KRAUS performed  better than $k$NN, although the difference was negligible.
Both methods worked better than PACE and KL20.
These results are summarized on boxplots of MSE in Figure~\ref{fig:imputing_mse}, as done already with the simulations based on Spanish temperatures.
The computational efficiency of $k$NN is reported in Table~\ref{tab:running_times}.

\section{Classification and outlier detection}

First we focus on the standard classification problem. Assume that each curve $X_i, 1 \leq i \leq n,$ comes from one of $G$ groups (subpopulations).
Let $Y_i$ be the group label of $X_i$. That is to say  $Y_i$ equals $y$ if $X_i$ comes from group $y$, $1\leq y\leq G$.
Given the \emph{training sample} $\{(X_i,Y_i)\}_{i=1}^n$  and a new curve $X$, the problem consists of predicting  the label $Y$ of $X$.
An \emph{ordinary} classifier is a rule that assigns to $X$ a group label $m(X)$.
Instead of outputting a group that $X$ should belong to,
a \emph{probabilistic} classifier is a prediction of the conditional probability distribution of $Y$.

\bigskip
There exists a wide variety of methods for classifying functional data \cite{WCM}.
Beyond treating the functional data as simple multivariate data in high dimensional spaces, 
many of these techniques make use of the fact that they are functions.
For example this is done by adding their derivatives, integrals, and/or other preprocessing functions to the analysis \cite{HRS17}.
The functional $k$NN classifier (f$k$NN) is a straightforward extension of the multivariate rule.
In a nutshell, one considers the $k$ nearest neighbors to the target curve and classifies it with the  more represented group.
This is the group to which the largest number of the $k$ nearest neighbors belong.
Then $k$ is chosen to minimize the empirical misclassification rate on the training sample.
The method introduced below is inspired by f$k$NN but differs from it in that  the  approach is probabilistic.

\subsection{The localization classifier}
\bigskip

Let $I_y$ be the set of indexes $i \in \{1,...,n\}$ for which $Y_i=y$.
Let $t_1,t_2,...,t_M$ be the time points at which the data are observed.
Although in practice they often come from a regular grid, 
in order to apply Theorem \ref{CLT01},  we assume that they are i.i.d. uniform random variables on  $[0,1]$.
Consider the empirical localization distance between $X$ and the group $y$. This is
\be
\label{ELD}
L_y(X) = \frac{1}{M} \sum_{r=1}^M L^{(k)} (X(t_r),  \{X_j(t_r),j \in I_y\} ).
\ee
Consider also its mean and variance 
\be
\label{muysigma}
 \mu_y=  \frac{1}{M} \sum_{r=1}^M \E \big[L^{(k)} (X(t_r), \{X_j(t_r),j \in I_y\} )] \ \ \mbox{and} \ \   \sigma_y^2 = \Var [L_y(X)].
\ee

Finally, consider the standardized score $\tau(X,y) = (L_y(X) -\mu_y)/\sigma_y$.
Denote by $T$ the random variable  $T =\tau(X,Y)$. 
We remark that $T$ not only depends on $(X,Y)$ but also on the training sample $\{(X_i,Y_i)\}_{i=1}^n$.
Assume the conditional distribution of $T$ given $\{Y=y\}$ is  absolutely continuous
with conditional probability density 
$f_y$  and  denote $\pi_y = \PP(Y= y)$. Then
the  Bayes rule implies
$$
\PP(Y=y|T) = \frac{\pi_y  f_y(T)}{\sum_{g=1}^G  \pi_g  f_g(T)}.
$$
Following the basic idea for functional discriminant analysis for classification  \cite{WCM}, we consider the  Bayes classifier
\begin{eqnarray}
\label{BayesC}
m^{(k)}(X) &= & \arg\max_y \PP(Y=y|T) \nonumber  \\
&=& \arg\max_y \pi_y  f_y(T). 
\end{eqnarray}
Theorem \ref{CLT01} implies that the conditional distribution of $T$ given $\{Y=y\}$ 
may be approximated by a standard normal distribution when $M$ is large.
Therefore, 
one might expect that the conditional probability density $f_y$ should be approximated by a standard normal density, denoted here by $\phi$.
If this is the case, we may consider the following approximation of  the 
Bayes classifier (\ref{BayesC}):
\begin{eqnarray}
\label{AsympB}
\tilde{m}^{(k)}(X) &=& \arg\max_y \pi_y  \phi(\tau(X,y)) \nonumber \\
&=& \arg\max_y \pi_y  \phi( \frac{L_y(X) -\mu_y}{\sigma_y}).
\end{eqnarray}
Here
we require knowledge of $\mu_y$ and $\sigma_y$. These values may be estimated from the training sample as follows.
First, consider the empirical localization distance between $X_i$ and its group $Y_i$. According to (\ref{ELD}),
this is
$$
L_i = \frac{1}{M} \sum_{r=1}^M L^{(k)} (X_i(t_r),  \{X_j(t_r), j \in I_{Y_i}\} ).
$$
Next, consider the sample mean and variance of the empirical localization distances for each group.
They are
\be
\label{samplemeanandvar}
\bar{L}_y= \frac{1}{n_y}\sum_{i=1}^n L_i  {\bf{1}}_{\{Y_i=y\}} \ \ \mbox{and} \ \  S_y^2 = \frac{1}{n_y-1}\sum_{i=1}^n (L_i - \bar{L}_y)^2  {\bf{1}}_{\{Y_i=y\}}.
\ee
These are empirical estimators of $\mu_y$  and $\sigma_y^2$ in (\ref{muysigma}).
Thus, by plugging these estimators into (\ref{AsympB}), we obtain the empirical classifier 
$$
\eta^{(k)}(X) = \arg\max_y \pi_y  \phi(  \frac{L_y(X) -\bar{L}_y} {S_y}).
$$
If $M$ is large and $n_y$ is  also large for any group label $y$, we expect that $\eta^{(k)}(X)$ is similar to the Bayes classifier $m^{(k)}(X)$.
Indeed, what we expect is
$$
\PP(Y=y|T)\approx \frac{\pi_y  \phi((L_y(X) -\bar{L}_y)/S_y)}{\sum_{g=1}^G  \pi_g  \phi((L_g(X) -\bar{L}_g)/S_g)}.
$$

\bigskip
We remark that, although the local feature of the empirical localization distances makes them robust in the presence of outliers, this is not the case of the sample mean and variance in (\ref{samplemeanandvar}).
The accuracy of these estimators may be clearly affected by the presence of outlier data.
For these reasons we consider trimmed means and variance, by discarding the outliers of each group, when calculating $\bar{L}_y$ and $S_y^2$ in (\ref{samplemeanandvar}).
The problem of outlier detection in a group,  say group  $y$,  is tackled by considering standard boxplots of  the samples of empirical localization distances
$\{L_i: Y_i = y\}$.
This tool is simple but powerful given the asymptotic Gaussianity of the empirical localization distances.
The problem of outlier detection in the complete population sample is addressed in a similar way by considering only one group.

\bigskip
If two or more groups are similar both in shape and scale, making  the classification difficult, then different $k$ values may provide different labels.
The same occurs when one applies f$k$NN: different nearest neighbors may belong to different groups.
In line with f$k$NN, we classify according to the more represented group.
In our case, we use the modal label  $\eta^{(1)}(X),\dots,\eta^{(k)}(X)$.
Also, as with f$k$NN, the $k$ value is chosen to minimize the empirical misclassification rate on the training sample.
We refer to this classification method by the {\em{ Localization classifier, or LC for short.}}

\bigskip
To evaluate the proposed methodology we performed a comparative study with benchmark methods for both outlier detection and classification.

\bigskip

\subsection{Classification study}

We compare f$k$NN and LC with two types of functional classifiers. 
On the one hand, we consider functional extensions of the Depth-to-Depth classifier (DD) \cite{LCL12} and on the  other hand, we consider the classifiers introduced by \cite{HRS17}, which also are inspired by DD but are based on special distances introduced by the cited authors instead of depths.
Both approaches map functions to points on the plane which require classifying by some bivariate classifier such as $k$NN.
The theory and methods for the former are discussed by \cite{cuesta2017} whereas   \cite{fdausc} developed the corresponding R package.
The R package related to the classifiers introduced by \cite{HRS17}  is also available at The Comprehensive R Archive Network \cite{mvpackage}.
Following these authors, we used $k$NN for bivariate classification;  both $k$NN and f$k$NN were  based on $L^2$-distances.
For each methodology we considered the three classifiers suggested by their authors, corresponding to different choices of depth and distance.
They are:
\begin{enumerate}
	\item DD$\_$hM, DD$\_$FM and DD$\_$MBD \cite{cuesta2017}.
	\item  fAO, fBD and  fSDO  \cite{HRS17}.
\end{enumerate}

\begin{figure}[!t]
	\includegraphics{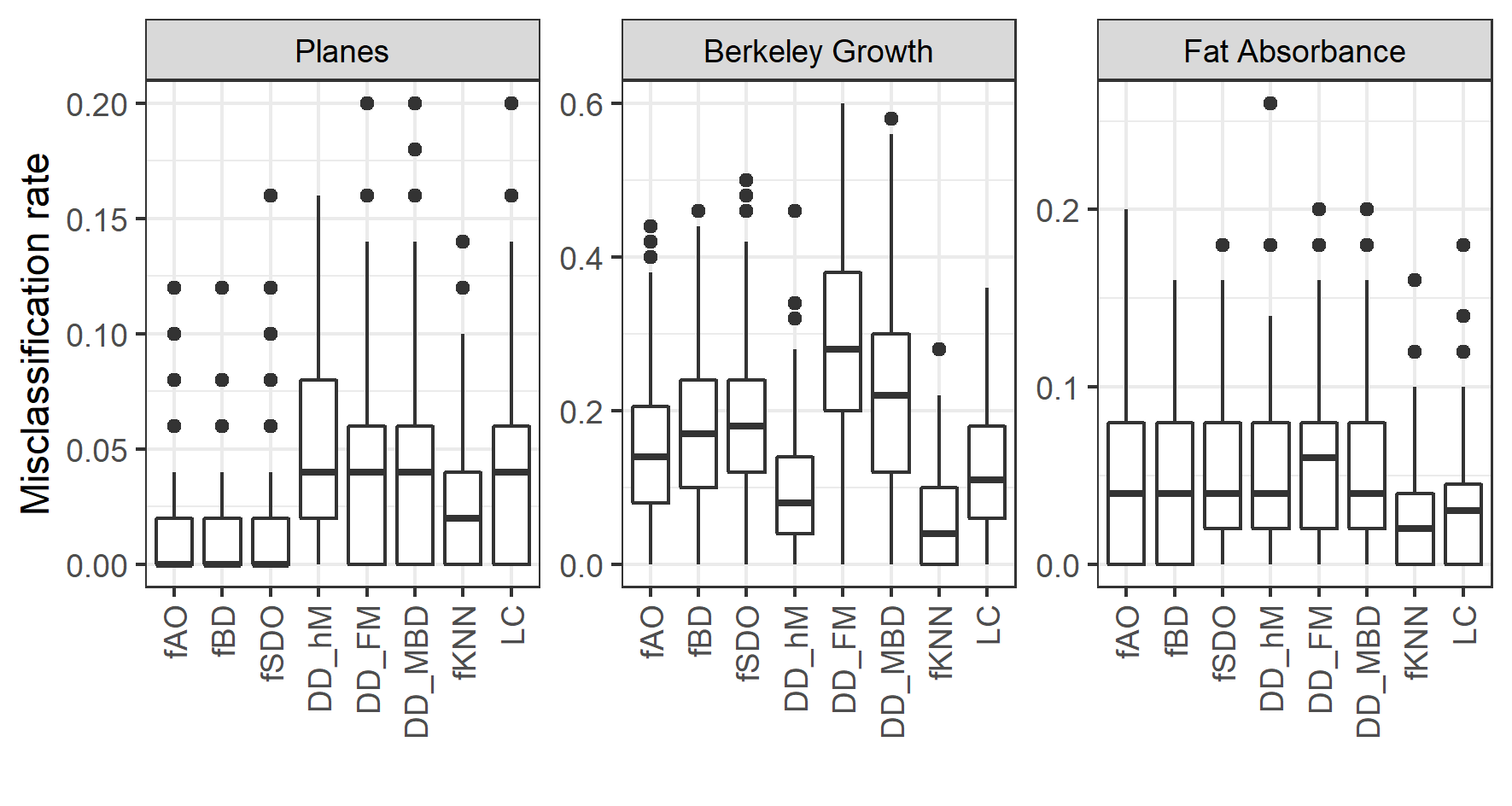}
	\caption{Misclassification rates from 1000 runs.}
	\label{fig:classification_results}
\end{figure}

\bigskip
We tested the eight methods under consideration on the following three examples considered in the literature to which we have referred:

\begin{enumerate}
	\item The fighter plane dataset used by  \cite{HRS17}. These are 210 univariate functions obtained from digital pictures of seven types of fighter planes, 30 from each type.
	\item Second derivative of fat absorbance from the \emph{Tecator} data set used by  \cite{cuesta2017}.
	For each piece of finely chopped meat we observe one spectrometric curve which corresponds to the absorbance measured at 100 wavelengths. The pieces are classified in  \cite{ferraty2006} into one of two classes according to small or large fat content. There are 12 pieces with low content and 103 with high.
	\item First derivative of the \emph{Berkeley growth study}. 
	This  dataset contains the heights of 39 boys and 54 girls from ages 1 to 18 and is a classic in the literature of FDA  \cite{ramsay:2005}. 
	\end{enumerate}
Complete descriptions of the datasets appear in  \cite{cuesta2017} and \cite{HRS17}. 
From each group of these datasets, we randomly selected one half of the data for
a training sample and we classified the rest.
We repeated this process one thousand times and reported the missed classification rates on boxplots in Figure~\ref{fig:classification_results}.

\bigskip
In summary, although all the methods perform well for the fighter plane dataset, the methods of \cite{HRS17}  were superior for these particular data.
The methods yielded larger misclassifications rates for the other two datasets, where f$k$NN was the best option, closely followed by LC.
Only DD$\_$hM was competitive with LC but at the cost of a long computational running time.
The pair ($k$NN, LC) offers a good classification tool, combining the efficient ordinary  classification by f$k$NN with the predicted probability provided by LC.

\subsection{Outlier detection}

One of the more popular tools for functional outlier detection is the functional boxplot  \cite{SG11}. 
This method mimics the univariate boxplot by  ordering the sample curves from the `median' outward according to the modified band depth \cite{sara:2009}.
It is well known that the functional boxplot detects magnitude outliers,
curves which  are outlying in part the observation domain.
However, the plot does not necessarily detect shape outliers.
They are sample functions that have different shapes from the bulk of data.
The outliergram \cite{AR14} and the MS-plot  \cite{Genton18, genton2019} were introduced for tackling both magnitude and shape outliers.

\begin{figure}[h!t]
	\includegraphics{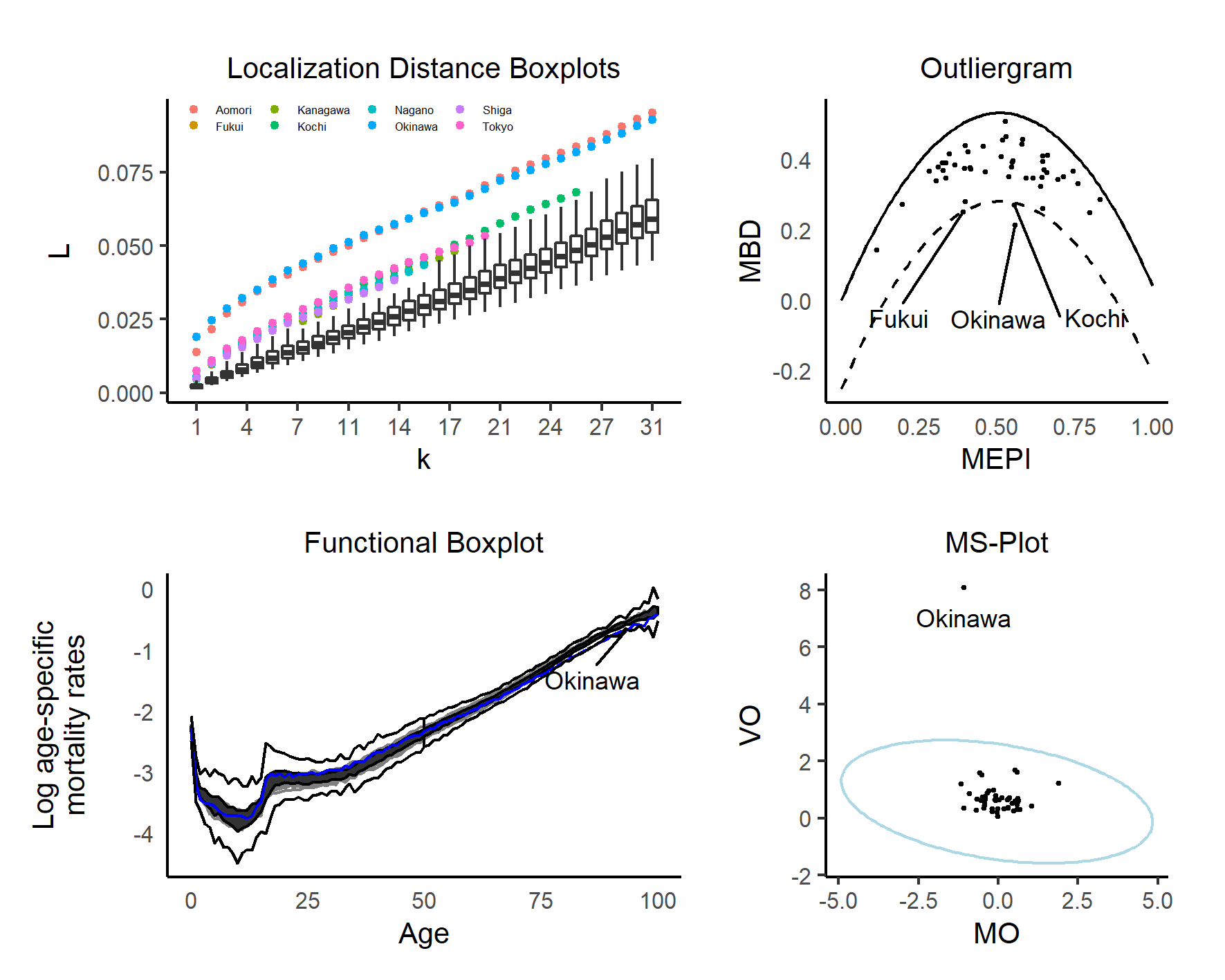}
	\caption{Outputs from the outlier detection methods under consideration. All the methods detect Okinawa. Outliergram and localization distance boxplots agree with respect to  Fukui and Kochi.  The localization distances are the only ones able to detect Aomori as an extreme outlier and the only ones indicating a certain atypicity of Tokyo. For a few values of $k$, the localization distances corresponding to Nagano, Shiga and Kanagawa fell above (but close to)  the default whiskers.}
	\label{fig:outliers_illustration1}
\end{figure}

\begin{figure}[h!]
	\includegraphics{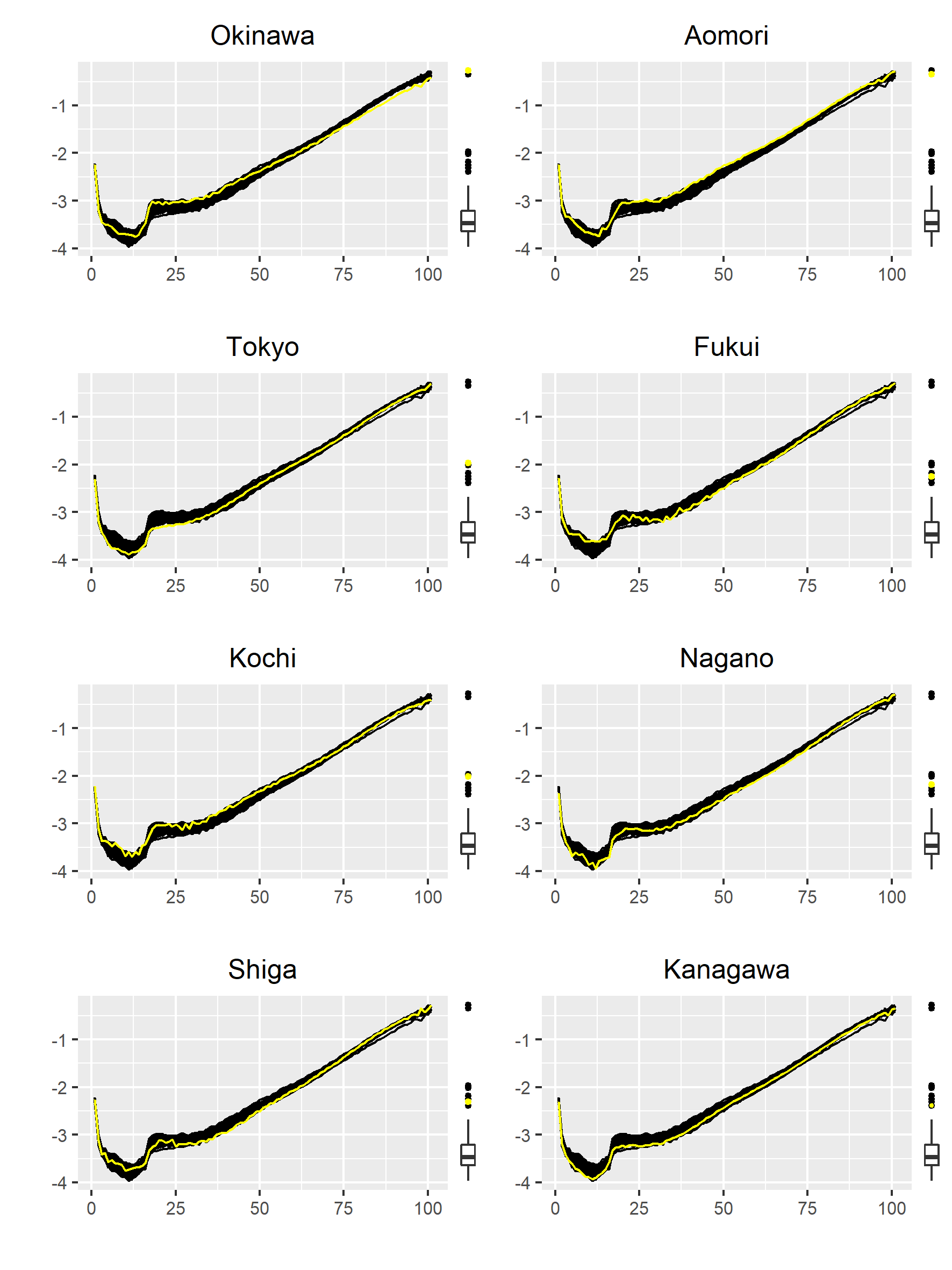}
	\caption{Log age-specific mortality rates and localization distances boxplot for $k=9$. Each outlier value and its corresponding curve are highlighted in yellow.}
	\label{fig:outliers_illustration2}
\end{figure}

\bigskip
We compare the above three methods with the method based on localization processes.
For this, we consider the Japanese mortality dataset discussed in Section 3.
For each of the $47$ prefectures we computed the average of the age-specific mortality rates between 1975 and 2007.
We only used data until 2007 because data from the Saitama prefecture is not available after this date.
In this way, we obtain 47 mortality curves smoothed by averaging over each prefecture.
Using the default parameters suggested by the authors, all the methods detected an outlier at Okinawa, where residents have famously lived longer than anywhere else in the world.
In addition, the outliergram also detected outliers at Fukui and Kochi.
Indeed, we observe that the curves corresponding to these two prefectures exhibit strong oscillations for ages below fifty years.
These oscillations are not seen in the greater part of the prefectures, which have smoother curves.
Regarding localization distances, the boxplots corresponding to Okinawa and Aomori were very extreme outliers for all the considered $k$ values.
We remark that, although the rest of the methods did not detect an outlier at Aomori, this prefecture has experienced the highest mortality rates for many years.
In fact, Aomori has been already considered an outlier by Japanese health officials \cite{Aomori}.
For several values of $k$, the localization distances corresponding to Fukui, Kochi and Tokyo fell above the default whiskers of the corresponding boxplots.
Indeed, we observe that Tokyo is a deep datum for ages above thirty years but it has extreme low values of mortality rates for ages below 25.
Finally, only for a few values of $k$, Nagano, Shiga and Kanagawa provide outliers with
respect to  localization processes and they fell close to the default whiskers.
In fact, by considering the more conservative upper whisker ($\mbox{Q}_3 + 3*\mbox{IQR}$ instead of $\mbox{Q}_3 + 1.5*\mbox{IQR}$) neither Tokyo, Nagano,  Shiga nor Kanagawa would be considered to be outliers.
However, it is interesting to observe that both Nagano and Shiga show shapes similar to Fukui and Kochi, although with more moderate oscillations.
Also, the behavior of Kanagawa is similar to Tokyo, but with less extreme mortality rates for ages below 25. The only value for which the eight mentioned prefectures fall above  the default whiskers is $k=9$.
All the above can be observed from Figures~\ref{fig:outliers_illustration1} and~\ref{fig:outliers_illustration2}.
In the former, we plot the outputs of all the methods under consideration.
In the latter, we show the particular case $k=9$, where the reader can inspect the curve of each prefecture under consideration.

\bigskip
In conclusion, though localization distance statistics agreed with the three benchmark methods with respect to  Okinawa and with the outliergram for Fukui and Kochi,  they recognize   Okinawa as an extreme outlier. In addition, the localization distances were able to detect Aomori as an extreme outlier and indicate a certain atypicality of Tokyo.
Also, they drew attention to a small departure from the bulk of prefectures of Nagano, Shiga and Kanagawa.
 
\section{Discussion}

The localization processes introduced here \eqref{lprocess}
are an alternative way to approximate curves from a given functional sample.
These processes  can be seen as  piecewise approximations  of different orders to a function from data collected in a functional setting.
Other estimation methods, for example those based on functional $k$NN and model-generated curves, often consider distances on function spaces for measuring nearness.
Unlike these methods, we consider the  localization width process formed by the point-by-point distances between the target curve  and its corresponding approximation.
Beyond the inherent interest of localization  processes, we introduce them  to provide a foundation for the rigorous asymptotic theory of nearest neighbor functional estimation.

\bigskip
First, we provide mean and distributional convergence of  localization widths when the number of sample curves increases up to infinity.
Under regularity conditions, we obtain $O(k/2n)$ bounds on the expected $L^1$ norm of the
difference between the $k$th localization process and the target. 
These results allow one to elucidate mild assumptions under which nearby neighbors to a target function on an observable range remain near  the target outside this range.
This property is the key to proving consistency of $k$NN type estimators for reconstructing curves from partially observed data. A particular $k$NN methodology is introduced and compared with three benchmark methods.
We present results of a simulation study based on yearly curves of daily Spanish temperatures and Japanese age-specific mortality rates,  two real world examples  where a large range of contiguous data is missing, but which may be reconstructed. 
Beyond the intuitive appeal of the method,
the results are  promising in terms of accuracy,  computational efficiency, and interpretability.


 \bigskip
Second,  the  central limit theorem  for empirical localization distances forms  the basis of new methods for classification as well as outlier detection.
These are problems where the $k$NN approach has been widely considered.
A comparative study  shows that the classification method proposed here is competitive with several benchmark methods.
Only $k$NN 
gave consistently superior results.
However, the new method predicts classification probabilities and provides standard normal scores that help validate the outputs rather than to only generate them. 
This makes  the method a useful complementary tool capable of providing probabilistic support to the ordinary $k$NN classification.
Regarding outlier detection, the case study considered shows that the method based on localization distances can detect both magnitude and shape outliers that other methods do not.

\bigskip
In conclusion,    the dual purpose of this paper has been to  introduce the $k$NN localization processes, the associated $k$NN localization distances, as well as mathematical tools lending rigor to both the current approach and to possible further approximation schemes based on nearest neighbors.  There remains the potential for further exploiting the asymptotic first and second order properties of  localization distances in functional data analysis.

\section{Proofs of main results in Section \ref{Sec2}}

\subsection{Auxiliary results}

We prepare for the proofs by first giving two lemmas. 

\begin{lemma} \label{Cox} For almost all $t \in [0,1]$, there are random variables $\{X'_j(t) \}_{j = 1}^n$, coupled to $\{X_j(t) \}_{j = 1}^n,$ and  a Cox process $\P_{\ka_t(X'_1(t))}$, also coupled to $\{X_j(t) \}_{j = 1}^n$,  such that
as $n \to \infty$
\begin{align*}
{W'_n}^{(k)}(t) & =
W^{(k)}(X'_1(t),  \{X'_j(t) \}_{j = 1}^n ) \nonumber \\
& =  \frac{2}  {k}   L^{(k)}(\0, n( \{X'_j(t) \}_{j = 1}^n  - X'_1(t)) )\\
&  \toP \frac{2}  {k}   L^{(k)}(\0, \P_{\ka_t(X'_1(t))}) =: {W'_{\infty}}^{(k)}(t).
\end{align*}
\end{lemma}

\noindent{\em Proof.}  The convergence may be deduced from Section 3 of  \cite{PY4} and we provide details as follows. 
For $x \in \R^d$ and $r > 0$ let $B(x,r)$ denote the Euclidean ball centered at $x$ with radius $r$.  Note that
$L^{(k)}$ is a stabilizing score function on Poisson input $\P$, that is to say that its value
at the origin is  determined by the local data consisting of the points in the 
intersection of the realization of $\P$ and the ball $B(\0, R^{L^{(k)}}(\0, \P))$,
where $R^{L^{(k)}}(\0, \P)$ is a radius of stabilization. 
For precise definitions we refer to \cite{PY4}, Section 3 and \cite{Pen07}.

The coupling of  Section 3 of \cite{PY4} shows that we may find 
$\{X'_j(t) \}_{j = 2}^n$, where $X'_j(t) \eqd X_j(t), j = 2,...,n$ and a Cox process
$ \P_{\ka_t(X'_1(t))}$ such that if we put
$\X'_{n - 1}(t) = \{X'_j(t) \}_{j = 2}^n$, then 
for all $K > 0$
\be \label{LemPY}
\lim_{n \to \infty} \PP(n(\X'_{n - 1}(t)  - X'_1(t)) \cap B(0,K) =
\P_{\ka_t(X'_1(t))} \cap B(0,K)) = 1.
\ee
See Lemma 3.1 of \cite{PY4}. 
Fix $\epsilon > 0$. 
Now write for all $\delta > 0$ 
\begin{align}
&\PP\left(  \left|\frac{2}  {k}   L^{(k)}(\0, n(\X'_{n - 1}(t)  - X'_1(t) ) - \frac{2}{k}   L^{(k)}(\0, \P_{\ka_t(X'_1(t))})\right| > \epsilon\right)\nonumber \\
& \leq \PP( n(\X'_{n - 1}(t)  - X'_1(t)) \cap B(0,K) \neq
\P_{\ka_t(X'_1(t))} \cap B(0,K)) \nonumber \\ 
& \hskip1.5cm + \PP( R^{L^{(k)}}(\0, \P_{\ka_t(X'_1(t))}) > K) \nonumber \\ 
& \leq \PP( n(\X'_{n - 1}(t)  - X'_1(t)) \cap B(0,K) \neq
\P_{\ka_t(X'_1(t))} \cap B(0,K)) \nonumber \\ 
& \hskip1.5cm + \PP( R^{L^{(k)}}(\0, \P_{\ka_t(X'_1(t))}) > K, \ka_t(X'_1(t)) \geq \delta)   + \PP(  \ka_t(X'_1(t)) \leq \delta).  \label{stab}
\end{align}
Given $\epsilon > 0$, the last term in \eqref{stab}
may be made less than $\epsilon/3$ if $\delta$ is small.  The penultimate term is bounded by
$\PP( R^{L^{(k)}}(\0, \P_{\delta}) > K)$, which is less than $\epsilon/3$ if $K$ is large, since 
$R^{L^{(k)}}(\0, \P_{\delta})$ is finite a.s. 
By \eqref{LemPY} the first term is  less than $\epsilon/3$ if $n$ is large.
Thus, for $\delta$ small and $K$ and $n$ large, the right-hand side of \eqref{stab} is less than 
$\epsilon$, which concludes the proof. \qed

\begin{lemma}  \label{boundedmom} Assume that the data is regular from 
below as at \eqref{reg}. 
Then $\sup_{n \leq \infty}\sup_{t \in [0,1]} \E W_n^{(k)}(t)^2 \leq C,$
where $C$ is a finite constant. 
\end{lemma}

\noindent{\em Proof.}  We treat the case $1 \leq n < \infty$, as the case
$n = \infty$ follows by similar methods.
Without loss of generality we assume
$S(\ka_t) = [0,1]$. We first prove the lemma for $k = 1$ and then for general $k$. 
Let $S_\delta$ be the subinterval  of $[0,1]$ such that $\ka_t(x) \geq \ka_{\text{min}}$
for all $x \in S_\delta$.
Note that $|S_\delta| =\delta \in (0,1]$ by assumption.
We have for all $r> 0$
\begin{align*}
\PP( W_n^{(1)}(X_1(t),  \{X_j(t) \}_{j = 2}^n ) \geq r) 
 & = \PP\left( L^{(1)}(X_1(t),  \{X_j(t) \}_{j = 2}^n ) \geq    \frac{r}{2n}\right)  \\
& = \Pi_{j = 2}^n \PP \left(|X_1(t) - X_j(t)| \geq \frac{r }{2n}\right) \\
& =  \left( 1 - \PP\left(|X_1(t) - X_2(t)| \leq  \frac{r }{2n}\right) \right)^{n - 1}\\
& = \left( 1 -    \int_{[0,1]} \int_{|x_1 - x_2| \leq  \frac{r}{2n}  } \ka_t(x_2)
 dx_2 \ka_t(x_1) dx_1   \right)^{n - 1}.
\end{align*}
For all $r \in (2n, \infty)$ we have
$\PP( L^{(1)}(X_1(t),  \{X_j(t) \}_{j = 2}^n ) \geq    \frac{r}{2n}) = 0$.
Thus we may assume without loss of generality that $r \in [0,2n]$.
We bound the double integral from below by
\begin{align*}
 \int_{[0,1]} \int_{|x_1 - x_2| \leq  \frac{r}{2n}  } \ka_t(x_2)
 dx_2 \ka_t(x_1) dx_1
 & \geq 
  \int_{S_\delta} \int_{|x_1 - x_2| \leq  \frac{r}{2n}  } \ka_t(x_2)
 dx_2 \ka_t(x_1) dx_1 \nonumber \\
& \geq 
  \int_{S_\delta} \int_{|x_1 - x_2| \leq  \frac{r \delta }{2n}  } \ka_t(x_2)
 dx_2 \ka_t(x_1) dx_1 \nonumber \\
& \geq \delta \ka_{\text{min}}^2 \cdot \frac{r \delta }{2n}.
\end{align*}
This gives 
\begin{align*}
\PP( W_n^{(1)}(X_1(t),  \{X_j(t) \}_{j = 1}^n ) \geq r)  & 
\leq \left( 1 -  \delta \ka_{\text{min}}^2 \cdot \frac{r \delta }{2n}  \right)^{n - 1}
\\
&
\leq c \exp{(- \frac{ \delta^2 \ka_{\text{min}}^2  r}{c}  )}, \ r > 0,
\end{align*}
where $c > 0$ is a constant.  Random variables having exponentially decaying tails have
finite  moments of all orders and thus this proves the lemma for $k = 1$.  

The proof for general $k$ uses the same approach.  
We show this holds for $k = 2$ as follows.  We have
$$
\PP( W_n^{(2)}(X_1(t),  \{X_j(t) \}_{j = 2}^n ) \geq r) 
= \PP\left( L^{(2)}(X_1(t),  \{X_j(t) \}_{j = 2}^n ) \geq    \frac{r}{n}\right).
$$
Given the event  $\{ L^{(2)}(X_1(t),  \{X_j(t) \}_{j = 2}^n ) \geq    \frac{r}{n} \}$,  either the first nearest neighbor to $X_1(t)$ is at a distance greater than 
$\frac{r  }{n}$ to $X_1(t)$ or the first nearest neighbor to $X_1(t)$ is at a distance less  than 
$\frac{r}{n}$ to $X_1(t)$ and the first nearest neighbor among the remaining $n - 2$ 
sample points exceeds $\frac{r}{n}$.

This gives
$$
\PP\left( L^{(2)}(X_1(t),  \{X_j(t) \}_{j = 2}^n ) \geq    \frac{r }{n}\right)  
\leq 
\PP\left( L^{(1)}(X_1(t),  \{X_j(t) \}_{j = 2}^n ) \geq    \frac{r}{n}\right) 
$$
$$
+ \sum_{i = 2}^n \PP\left( L^{(1)}(X_1(t),  \{X_j(t) \}_{j = 2, j \neq i}^n ) \geq    \frac{r }{n}\right) \PP( |X_1 - X_i| \leq  \frac{r }{n}).
$$
Since $\PP( |X_1 - X_i| \leq  \frac{r }{n}) = O(n^{-1})$ for all $i = 2,...,n$  we may use
 the bounds for the case 
$k = 1$ to show that 
$\PP\left( L^{(2)}(X_1(t),  \{X_j(t) \}_{j = 2}^n ) \geq    \frac{r}{n}\right)  $
decays exponentially fast in $r$.  The proof for general $k$ follows in a similar fashion and we leave the details to the reader.   This proves the lemma. 
 \qed

\subsection{ Proof of Theorem \ref{mainthm1}}

We first prove \eqref{WLLN}. 
Let $t \in [0,1]$ be such that the marginal density $\kappa_t$ exists.
By translation invariance of $L^{(k)}$ we have as $n \to \infty$
\begin{align*}
\E W_n^{(k)}(t)
& = \E  L^{(k)}\Big( \frac{2n} {k  }  X_1(t),  \frac{2n} {k } \{X_j(t) \}_{j = 1}^n \Big)  \\
& = \frac{2} {k  } \E L^{(k)}(\0,  n( \{X_j(t) \}_{j = 1}^n  -  X_1(t) )  )  \\
& \to \int_{  S(\kappa_t)}  \frac{2}  {k}  \E L^{(k)} (\0, \P_{\kappa_t(y)}  ) \kappa_t(y) dy,
\end{align*}  
where the limit  follows since  convergence in probability (Lemma 
\ref{Cox})  combined with uniform integrability (Lemma \ref{boundedmom})
gives convergence in mean. 
  Now for any constant $\tau$ we have
$\E  L^{(k)} (\0, \P_{\tau }  ) = \tau^{-1}  \E L^{(k)} (\0, \P_{1}  )$. Notice that  $ L^{(k)} (\0, \P_{1}  )$ is a Gamma $\Gamma(k,2)$ random variable with shape parameter $k$ and scale parameter $2$ and thus $\E L^{(k)} (\0, \P_{1}) = k/2$.  The proof of  \eqref{WLLN} is complete.

\bigskip

To prove \eqref{Var}, we replace $W_n^{(k)}(t)$ by its square in the above computation.  This yields 
\begin{align*}
\lim_{n \to \infty} \E W_n^{(k)}(t)^2
= \int_{  S(\kappa_t)} \frac{4}  {k^2} \E (L^{(k)} (\0, \P_{\kappa_t(y)}  ))^2 \kappa_t(y) dy.
\end{align*} 
For any constant $\tau \in (0, \infty)$ we have
\be \label{Gamma}
\E (L^{(k)} (\0, \P_{\tau }  ))^2 = \tau^{-2}  \E ( L^{(k)} (\0, \P_{1}  ))^2 = 
\tau^{-2} \frac{(k + 1)k} {4},
\ee
 since  the second moment of a Gamma $\Gamma(k,2)$ random 
 variable equals $(k + 1)k/4$.
These facts yield \eqref{Var}. 
The limit \eqref{WLLN2} is a consequence of Lemma \ref{Cox}.
\qed

\subsection{ Proof of Theorem \ref{global}}
\bigskip
By Lemmas \ref{Cox} and \ref{boundedmom}, the random variables
${W'_n}^{(k)}(t)  =
W^{(k)}(X'_1(t),  \{X'_j(t) \}_{j = 1}^n )$  converge in probability and also in mean.
It follows that  as $n \to \infty$ 
$$
F_n(t) = \E |   {W'_n}^{(k)}(t) - {W'_{\infty}}^{(k)}(t)|
= \E | {W}_n^{(k)}(t) - {W}_{\infty}^{(k)}(t)| \to 0.
 $$
 Now $\sup_n \sup_t F_n(t) \leq C$ and the  bounded convergence theorem gives
 $\lim_{n \to \infty} \int_0^1  F_n(t) dt = 0.$  This gives the first statement of 
Theorem \ref{global}.  The identity $\lim_{n \to \infty} \int_0^1 \E {W}_{\infty}^{(k)}(t) dt = 1$ follows from
$$
 \int_0^1  F_n(t) dt \geq \left|  \int_0^1 \E  W_n^{(k)}(t)dt
 -  \int_0^1 \E {W}_{\infty}^{(k)}(t) dt \right|
 $$
 and the identity $\E {W}_{\infty}^{(k)}(t) =  |S(\kappa_t)|.$  \qed

\subsection{ Proof of Theorem \ref{mainthm2}}

\noindent   Lemma \ref{Cox}  assumes that $k$ is fixed.  The lemma  will not always hold if
$k$ is growing with $n$.  Thus our proof techniques and coupling arguments  need to be modified. We break the proof of Theorem \ref{mainthm2} into five parts. 

\vskip.3cm

\noindent{\em Part (i)  Coupling}. We start with a general coupling fact.  Given Poisson point processes $\Sigma_1$ and $\Sigma_2$ with densities $f_1$ and $f_2$, we may find coupled  
 Poisson point processes $\Sigma'_1$ and $\Sigma'_2$ with 
 $\Sigma'_1 \eqd  \Sigma_1$ and  $\Sigma'_2 \eqd  \Sigma_2$ such that the probability that the two point processes are not equal on $[-A,A]$ is
 bounded by
 $$
 \int_{-A}^{A} |f_1(x) - f_2(x) | dx.
 $$
 
 Let $t \in [0,1]$ be such that the marginal density $\kappa_t$ exists. 
   As in Theorem \ref{mainthm2}, we assume that $\kappa_t$ is  $\alpha$-H\"older continuous for  $\alpha \in (0,1]$.
 Note that the point process
$n(\P_{n \kappa_t} - y)$ has intensity density $\ka_t( \frac{x} {n} + y), \ x \in  n   (S(\kappa_t)  - y)$.      
For each $y \in S(\kappa_t)$, we  may find coupled Poisson point processes $\P'_{n \kappa_t}$ and $ \P'_{\kappa_t(y)}$ with 
 $ n(\P'_{n \kappa_t} - y) \eqd  n(\P_{n \kappa_t} - y)$ and  $\P'_{\kappa_t(y)} \eqd \P_{\kappa_t(y)}$
such that the probability that the point processes
 $ n(\P'_{n \kappa_t} - y)$ and $\P'_{\kappa_t(y)}$ 
 are not equal on $[-A, A]$ is bounded uniformly in $y \in S(\ka_t)$ by
\be \label{coupling}
\int_{-A}^{A} | \ka_t\left( \frac{x} {n} + y\right) - \ka_t(y) | dx \leq 2A \left(\frac{ A} {n}\right)^{\alpha}.
\ee
We will need this coupling in what follows.

\vskip.3cm

\noindent{\em Part (ii)}  {\em Poissonization.}  We will first show a Poissonized version of 
\eqref{WLLNk}. 
Write $k$ instead of $k(n)$.
 We assume that we are given a Poisson number of data curves $\{X_j(t) \}_{j = 1}^{N(n)}$, where $N(n)$ is an independent Poisson random variable with parameter $n$. We aim to show 
\be \label{WLLNkPoisson}
\lim_{n \to \infty}  \E W_n^{(k)}(X_1(t),  \{X_j(t) \}_{j = 1}^{N(n)} )  =  |S(\kappa_t)|.
\ee

By translation invariance of $W_n^{(k)}$ we have 
\begin{align}
\E W_n^{(k)}(X_1(t),  \{X_j(t) \}_{j = 1}^{N(n)} )
& = \frac{2n} {k}  \E L^{(k)}\left(   X_1(t),   \{X_j(t) \}_{j = 1}^{N(n)} \right) \nonumber  \\
& = \frac{2n} {k} \E L^{(k)}(\0,  ( \{X_j(t) \}_{j = 1}^{N(n)} - X_1(t) )  ). \label{A}
\end{align}

We assert that as $n \to \infty$ 
\be \label{assert}
\left| \E L^{(k)}(\0,  \frac{2n} {k} ( \{X_j(t) \}_{j = 1}^{N(n)} - X_1(t) )  )
 -  \frac{2} {k}  \E  L^{(k)}(\0, \P_{\ka_t(X_1(t))}) \right| \to 0.
 \ee
  Combining \eqref{A}-\eqref{assert} and recalling that  
$\frac{2} {k} \E  L^{(k)}(\0, \P_{\ka_t(X_1(y))}) =  |S(\kappa_t)|,$ we obtain
  $$ \lim_{n \to \infty}   \E W_n^{(k)}(X_1(t),  \{X_j(t) \}_{j = 1}^{N(n)} )  =  |S(\kappa_t)|,
 $$
which establishes \eqref{WLLNkPoisson}. 
 
 It remains to establish \eqref{assert}. 
 A Poisson point process on a set $S$  with intensity density $n h(x)$, where $h$ is itself a density, may be expressed as the realization of random variables $X_1,....,X_{N(n)}$, where $N(n)$ is 
 an independent Poisson random variable with parameter $n$ and where each $X_i$ has density $h$
 on $S$. 
 Thus the point process $\{X_j(t) \}_{j = 1}^{N(n)}$ is the Poisson point process
$\P_{n\ka_t}$.  To show the assertion \eqref{assert}
 we thus need to show
$$
\left| \frac{2} {k} \E L^{(k)}(\0, n(\P_{n\ka_t} - X_1(t)))  - \frac{2} {k} 
\E  L^{(k)}(\0, \P_{\ka_t(X_1(t))}) \right| \to 0
$$
or equivalently,
$$
\left| \frac{1} {k} \E L^{(k)}(\0, n(\P'_{n \ka_t} - X'_1(t)))  - \frac{1} {k} 
\E  L^{(k)}(\0, \P'_{\ka_t(X'_1(t))}) \right| \to 0,
$$
where $\P'_{n \kappa_t}$ and  $\P'_{\kappa_t(y)}$ are as in part (i). 

Fix $\epsilon > 0.$  As in the bound \eqref{stab}, we have  
for all $\delta > 0, K > 0$ 
\begin{align}
&
\PP\left(  \left|\frac{2} {k} L^{(k)}(\0, n(\P'_{n\ka_t} - X'_1(t))) - \frac{2}{k}   L^{(k)}(\0, \P'_{\ka_t(X'_1(t))})\right| > \epsilon \right)   \nonumber \\
&
 \leq \PP\left( \frac{2n} {k} (\P'_{n\ka_t} - X'_1(t))  \cap B(0,K) \neq
\frac{2} {k} \P'_{\ka_t(X'_1(t))} \cap B(0,K) \right) \nonumber \\ 
&
\hskip1.5cm + \PP\left( R^{L^{(k)}}(\0, \frac{2} {k} \P'_{\ka_t(X'_1(t))}) > K\right) \nonumber \\ 
&
\leq \PP\left( \frac{2n} {k} (\P'_{n\ka_t} - X'_1(t)) \cap B(0,K) \neq
 \frac{2} {k} \P'_{\ka_t(X'_1(t))} \cap B(0,K)\right) \nonumber \\ 
&
 \hskip1.5cm + \PP\left( R^{L^{(k)}}(\0, \frac{2} {k} \P'_{\ka_t(X'_1(t))}) > K, \ka_t(X'_1(t)) \geq \delta\right)   + \PP(  \ka_t(X'_1(t)) \leq \delta).  \label{stab1}
\end{align}

The last term in \eqref{stab1}
may be made less than $\epsilon/3$ if $\delta$ is small.  The penultimate term is
 bounded 
$\PP( R^{L^{(k)}}(\0, \P_{\delta k}) > K) = \PP( R^{L^{(k)}}(\0, \frac{1}{\delta k} \P_{1}) > K)
= \PP( \frac{1}{\delta k} \Gamma(k,2) >K)$, 
which by Chebyshev's inequality is less than $\epsilon/3$ if $K$ is large. 
The first term satisfies
\begin{align}
&
 \PP\left( \frac{2n} {k} (\P'_{n \ka_t} - X'_1(t))  \cap B(0,K) \neq
\frac{2} {k} \P'_{\ka_t(X'_1(t))} \cap B(0,K) \right)
\nonumber \\
&
=  \PP\left( n (\P'_{n \ka_t} - X'_1(t))  \cap B(0, \frac{Kk}{2} ) \neq
 \P'_{\ka_t(X'_1(t))} \cap B(0, \frac{Kk}{2}) \right)
\nonumber \\
& \leq  Kk (  \frac{Kk}{2n})^{\alpha}
\end{align}
where the inequality follows from \eqref{coupling}. 
By assumption, we have   $\lim_{n \to \infty} \frac{ k^{1 + \alpha}}{n^{\alpha}} = 0$  and it follows that the first term is  less than $\epsilon/3$ if $n$ is large.
Thus, for $\delta$ small and $K$ and $n$ large, the right-hand side of \eqref{stab} is less than 
$\epsilon$. Thus
$$
\left|\frac{2} {k} L^{(k)}(\0, n(\P'_{\ka_t} - X'_1(t))) - \frac{2}{k}   L^{(k)}(\0, \P'_{\ka_t(X'_1(t))})\right| \toP 0.
$$
The assertion \eqref{assert} follows since convergence in probability combined with 
uniform integrability gives convergence in mean.

\vskip.3cm 
\noindent{\em Part (iii)  de-Poissonization.}   We  de-Poissonize the above  
equality  to obtain \eqref{WLLNk}. In other words we need to show that the limit
does not change when $N(n)$ is replaced by $n$.  
Put
$$
 {\cal Y}_n = \begin{cases} X_1(t),...,X_{N(n)- (N(n) - n)^+}(t)
 , & {\rm if} \ N(n) \geq n
\\
X_1(t),...,X_{N(n) + (n  - N(n)^+}(t)
 , & {\rm if} \ N(n) < n.
\end{cases}
$$
Then ${\cal Y}_n \eqd \{ X_1(t), X_2(t),...,X_{n}(t)\}.$  We use this coupling
of Poisson and binomial input  in all that follows.

We wish to show that $\hat{X}_{n,1}^{(k)}(t)$ coincides with 
$\hat{X}_{N(n),1}^{(k)}(t)$ on a high probability event; in other words we wish to show that the
sample points with indices between $\min(n, N(n))$ and 
$\max(n, N(n))$ do not, in general, modify the value of $\hat{X}_{n,1}^{(k)}(t)$.
Consider the event that the Poisson random variable does not differ too much 
from its mean, i.e., 
$$
E_n= \{|N(n) - n| \leq c \sqrt{n} \log n \}
$$
and note that tail bounds for Poisson random variables show that
there is $c > 0$ such that   $\PP(E_n^c) = O( n^{-2})$. 
Write
\begin{align*}
& \E |  W_n^{(k)}(X_1(t),  \{X_j(t) \}_{j = 1}^{N(n)} ) - W_n(t) | \\
& \leq \E  |  [W_n^{(k)}(X_1(t),  \{X_j(t) \}_{j = 1}^{N(n)} ) - W_n^{(k)}(t)]
{\bf 1}(E_n)  | \\
& \hskip1cm + \E  |  [W_n^{(k)}(X_1(t),  \{X_j(t) \}_{j = 1}^{N(n)} ) - W_n^{(k)}(t)]
{\bf 1}(E_n^c) |.
\end{align*}
The last summand is $o(1)$, which may be seen using the  Cauchy-Schwarz inequality,  Lemma \ref{boundedmom}, and $\PP(E_n^c) = O(n^{-2})$. 

For any $1 \leq j \leq c \sqrt{n} \log n$ we define
$$
A_{n,j} = A_{n,j}(t) =  \{|X_{\min(n, N(n)) + j}(t) - X_1(t) | \geq | \hat{X}_{n,1}^{(k)}(t)
- X_1(t) | \}.
$$
This is the event that  the data curves  having  index  larger than 
$\min(n, N(n)) $ are farther away from  $X_1(t)$ than is the data curve  $\hat{X}_{n,1}^{(k)}(t)$. 

Given i.i.d. random variables $Z_i, 1 \leq i \leq n$,  we let 
$Z_i^{(k)}$ denote the $k$th nearest neighbor to $Z_i$. 
 Given an independent random variable $Z_0$ having the same distribution as $Z_i$, the probability that
$Z_0$ belongs to $[Z_i,  Z_i^{(k)}]$ coincides with the probability that a uniform random variable
on $[0,1]$ belongs to $[U_i,  U_i^{(k)}]$  where 
$U_i, 1 \leq i \leq n,$ are  i.i.d. uniform random variables
on $[0,1].$ By exchangeability this last probability equals $k/(n-1)$.

It follows that for any $j = 1,2,\dots$
$$
\PP(A_{n,j}| n \leq N(n) )  =  \frac{(n - 1) - k} { n -1 },
$$
whereas 
$$
\PP(A_{n,j}| n \geq N(n) )  =  \frac{(N(n) - 1) - k} {N(n) -1 }.
$$

We have
$$
\PP(A_{n,j}| N(n))  =  1 - \frac{k} { \min((n -1), (N(n) - 1)) }
$$
and thus
$\PP(A_{n,j} \cap E_n)  \geq 1 - \frac{k} { (n - 1) - c \sqrt{n} \log n}.$
Thus
\begin{align*}
& \PP((W_n^{(k)}(X_1(t),  \{X_j(t) \}_{j = 1}^{N(n)} ) - W_n^{(k)}(t)) {\bf{1}}(E_n) 
 \neq 0 )  \\
 & 
 = 1 - \left(\PP(A_{n,1} \cap E_n)\right)^{c \sqrt{n} \log n} \\
&  \leq 1 - \left( 1 - \frac{k} { (n - 1) - c \sqrt{n} \log n }\right)^{c \sqrt{n} \log n}\\
& \leq \frac{k c' \log n} { \sqrt{n}  }.
\end{align*}
When $\frac{k c' \log n} { \sqrt{n}  } = o(1)$ we find that 
$\left(W_n^{(k)}(X_1(t),  \{X_j(t) \}_{j = 1}^{N(n)} ) - W_n^{(k)}(t)\right) {\bf{1}}(E_n)$
converges to zero in probability as $n \to \infty$, and thus so does 
$\left(W_n^{(k)}(X_1(t),  \{X_j(t) \}_{j = 1}^{N(n)} ) - W_n^{(k)}(t)\right).$
By uniform integrability we obtain that 
$(W_n^{(k)}(X_1(t),  \{X_j(t) \}_{j = 1}^{N(n)} ) - W_n^{(k)}(t))$ converges to zero in mean.  This completes the proof of  \eqref{WLLNk}.

\bigskip
\noindent{\em Part (iv) Variance convergence. }
Replacing  $W_n^{(k)}(X_1(t),  \{X_j(t) \}_{j = 1}^{N(n)})$ by its square in the above computation gives
 \begin{align*} 
 & \lim_{n \to \infty}[ \E W_n^{(k)}(X_1(t),  \{X_j(t) \}_{j = 1}^{N(n)} )^2  - (\E W_n^{(k)}X_1(t),  \{X_j(t) \}_{j = 1}^{N(n)} )^2] \\
 & =  \lim_{n \to \infty}  \frac{4} {k^2 }  \int_{ S(\kappa_t)    }    \E  L^{(k)}(\0, \P_{\ka_t(y)})^2  \ka_t(y) dy  - |S(\kappa_t)|^2\\
 & = \int_{S(\kappa_t)} \frac{  1 } {\ka_t(y)} dy - |S(\kappa_t)|^2,
  \end{align*}
where the last equality makes use of \eqref{Gamma}.    This gives
 \eqref{Vark}, as desired. 
\bigskip

\noindent{\em Part (v)  $L^1$  convergence. }
The limit \eqref{WLLNkLone} follows exactly as in the proof of Theorem \ref{global}.  
\qed

 \bigskip

 \subsection{ Proof of Theorem \ref{CLT01}}
 
This result is a straightforward consequence of the  central limit theorem for $M$-dependent random variables. It is enough to prove the central limit theorem for the 
re-scaled random variables $\{L(mT_r)\}_{r = 1}^m$.
 Indeed these 
 random variables $\{L(mT_r)\}_{r = 1}^m$ have moments of all orders and they are
 $M$-dependent since $\{L(mT_r)\}_{r \in A}$ and $\{L(mT_r)\}_{r \in B}$ 
 are independent
 whenever the distance between the index sets $A$ and $B$ exceeds $2M$. The asserted asymptotic normality follows by the classical central limit theorem for $M$-dependent random variables.  \qed
 
 \section{Appendix }

Here we discuss asymptotic normality of average distances for large $n$. 
When functional data are partially observed,  some sample functions are observed at time $t$ whereas others are not \cite{KL20, Kraus15, YMW12}.
For modeling this setting, we suppose we have a total of $N(n)$ curves, $N(n)$ being a Poisson random variable with mean $n$, and a proportion of them, say $pN(n)$, are not observed at $t$.
Therefore, only $(1-p)N(n)$ are available for computing the localization process at $t$.

\bigskip
Formally, for each $t \in  [0,1]$, we let $X_i(t), 1 \leq i \leq N(n)$, be a marked point process with values in $\R \times \M$, where the mark space $\M= \{0,1\}$ is equipped with a measure $\mu_{\M}$  giving probability $p$ to $\{1\}$ and probability
 $1-p$ to $\{0\}$. If the mark at $X_i(t)$  equals one, then it comes from a curve whose value is unknown at time $t$, 
 whereas if a mark at a point equals zero, then it comes from the collection of curves whose values are known at time $t$ and which are thus 
 used to construct the localization process at $t$. In general, the marks for  $X_i(t_1)$ and $X_i(t_2)$ are different for $t_1 \neq  t_2$. 
We write $\tilde X_i(t)$ to denote the point $ X_i(t)$ equipped with a mark.
Let $I_m(t), m \in \{0,1\},$ be the set of indices $i$'s for which  $\tilde X_i(t)$ has mark equal to $m$.
This gives the statistic
$$
W_n^{(k)}\Big(\tilde X_i(t),  \{\tilde X_j(t)\}_{j=1}^{N(n)}\Big) = W_n^{(k)}(X_i(t),  \{ X_j(t), j \in I_0(t)\} )
$$
defined as the  re-scaled distance \eqref{width} but only based on the available (observed) points at $t$.

\bigskip
To develop the second order limit theory,
we write $\tilde x$ to denote the point $x$ equipped with a mark.
We write $\tilde \P_{\ka_t(y)}$ for the Poisson point process $\P_{\ka_t(y)}$ where each point is equipped with an independent mark having measure $\mu_{\M}$, and we put
\begin{align}  \label{defmarkedvar}
\tilde \sigma^2_t & = \int_{  S(\kappa_t)  } \E (  L^{(k)}(\tilde \0, \tilde \P_{\ka_t(y)}) )^2  \ka_t(y) dy \\
& \ \ \ \ \ \ \ \ + \int_{ S(\kappa_t)   } \int_{\tilde x \in \tilde \R}  [ \E ( L^{(k)}(\tilde \0, \tilde \P_{\ka_t(y)} \cup \{\tilde x \} )  L^{(k)}(\tilde x, \tilde \P_{\ka_t(y)} \cup \{ \tilde \0\} )
\nonumber \\
& \ \ \ \ \ \ \ \ \ \ \ \ \ \ - \E (  L^{(k)}(\tilde \0, \tilde \P_{\ka_t(y)}  ) \E L^{(k)}(\tilde x, \tilde \P_{\ka_t(y)}) )] (\ka_t(y))^2 d \tilde x dy.\nonumber
\end{align}

\bigskip
Let $d_K(X,Y)$ be the Kolmogorov distance between random variables $X$ and $Y$ and let   $N(\mu,\sigma^2)$ denote a normal random variable centered at $\mu$ with  variance equal to $\sigma^2$. 

\begin{theorem}  \label{CLT}
We assume that $\ka_t$ is Lipschitz and bounded away from zero on its support  $S(\kappa_t)$.
 We have for $p\in [0,1)$ and $k \in \N$
\be \label{markedCLTfixt}
\frac{  \sum_{i \in I_0(t)} \Big(W_n^{(k)}\Big(\tilde X_i(t),  \{\tilde X_j(t) \}_{j = 1}^{N(n)}\Big) - \frac{|S(\kappa_t)| } {1 - p} \Big)} { \sqrt{n(1-p)}} \tod  N \bigg(0, \nu^2(t,k) \bigg)
\ee
where
\be \label{markedVart}
\nu^2(t,k) = \lim_{n \to \infty} \frac{ \Var  \sum_{i \in I_0(t)}  W_n^{(k)}\Big(\tilde X_i(t),  \{\tilde X_j(t) \}_{j = 1}^{N(n)} \Big) } { n(1-p)  } =  \frac{4 \tilde \sigma_t^2}{(1-p)^2k^2}
\ee
and where $\tilde \sigma_t^2$ is at \eqref{defmarkedvar}.
Moreover for all $k \in \N$ there is a constant $C_1(k)$ such that
\be \label{markedCLTrate}
d_K \left( \frac{  \sum_{i \in I_0(t)} \Big( W_n^{(k)}\Big(\tilde X_i(t),  \{\tilde X_j(t) \}_{j = 1}^{N(n)} \Big) - \frac{|S(\kappa_t)| } {1 - p} \Big)}
 { \sqrt{ \Var  \sum_{i \in I_0(t)}  W_n^{(k)}\Big(\tilde X_i(t),  \{\tilde X_j(t) \}_{j = 1}^{N(n)} \Big)   }},  N(0,1) \right) \leq \frac{ C_1(k) } {\sqrt {n}}.
\ee
Also we have for $p\in (0,1)$
\be \label{markedCLTrate2}
d_K \left( \frac{  \sum_{i \in I_1(t)} \Big( W_n^{(k)}\Big(\tilde X_i(t),  \{\tilde X_j(t) \}_{j = 1}^{N(n)} \Big) - \frac{|S(\kappa_t)| } {1 - p} \Big)}
 { \sqrt{ \Var  \sum_{i \in I_1(t)}  W_n^{(k)}\Big(\tilde X_i(t),  \{\tilde X_j(t) \}_{j = 1}^{N(n)} \Big)   }},  N(0,1) \right) \leq \frac{ C_1(k) } {\sqrt {n}}
\ee
and
\be \label{markedratio}
\lim_{n \to \infty} \frac{ \Var  \sum_{i \in I_1(t)}  W_n^{(k)}\Big(\tilde X_i(t),  \{\tilde X_j(t) \}_{j = 1}^{N(n)} \Big) } { \Var  \sum_{i \in I_0(t)}  W_n^{(k)}\Big(\tilde X_i(t),  \{\tilde X_j(t) \}_{j = 1}^{N(n)} \Big)  } =  \frac{p}{1-p}.
\ee
\end{theorem}
 
 \bigskip
Theorem \ref{CLT} provides asymptotic confidence intervals for the average error when replacing   unobserved curves with their corresponding localization processes. 
This average error is
\begin{eqnarray}
\label{averageerror}
 \bar{L}^{(k)}(t) &=&   \frac{1}{\mbox{card}[I_1(t)]}\sum_{i \in I_1(t)} L_n^{(k)}\Big(\tilde X_i(t),  \{\tilde X_j(t) \}_{j = 1}^{N(n)} \Big)\\
 & = &  \frac{1}{\mbox{card}[I_1(t)] }\ 
 \frac{k}{2n}\sum_{i \in I_1(t)} W_n^{(k)}\Big(\tilde X_i(t),  \{\tilde X_j(t) \}_{j = 1}^{N(n)} \Big),\nonumber
\end{eqnarray}
with $\mbox{card}[A]$ being the cardinality of $A$.
In fact, Theorem \ref{CLT} and Slutsky's theorem imply that, if $n$ is large enough, the distribution of
$\bar{L}^{(k)}(t)$  is approximately normal with mean equal to $(k/2n)\cdot 1/(1-p)$ and variance equal to $(k/2n)^2\cdot \nu^2(t,k)/np$.
We observe that these values are, respectively, $O(k/n)$ and $O(k^2/n^3)$.
To verify the latter, we require that $\tilde \sigma^2_t$  at (\ref{defmarkedvar}) increases as $k^2$.
This can be seen from the proof of \eqref{Var}.

\bigskip
Leaving aside the statistical applications that involve partially observed functional data, we remark that Theorem \ref{CLT} also establishes a central limit theorem for the sum of the localization distances when all the curves are completely observed. This corresponds to the case $p=0$.  It shows, given a Poisson point process on the interval  $[0,1]$ with intensity density $n \ka_t$,  that the sum of the re-scaled distances between points of this point process and their $k$th nearest neighbors is asymptotically normal as $n \to \infty$.  The case $p = 0$ is a special case of a more general result of \cite{PY1}
giving the total edge length of the $k$ nearest neighbors graph on Poisson input in all dimensions when the entirety of the input is used.

 \subsection{ Proof of Theorem \ref{CLT}}

The next result gives a rate of convergence of mean distances.  While it is of independent interest, we will use it in the proof of Theorem \ref{CLT}.   

\begin{prop} (rate of convergence of expected width on Poisson input and marked Poisson input) Assume $t \in [0,1]$ is such that $\ka_t$ exists and satisfies the conditions of Theorem  \ref{CLT}.
For all $k = 1,2,...$ there is a constant $c(k)$ such that 
\be \label{rate}
\left|  \E W_n^{(k)}(X_1(t),  \{X_j(t) \}_{j = 1}^{N(n)} )  -  |S(\kappa_t)|  \right| \leq \frac{c(k)} {n},  \ n \geq 1.
\ee
 Also,  on the event $\{ i \in I_0(t) \}$ we have 
\be \label{rateA}
\left|  \E W_n^{(k)}( \tilde{X}_i(t),  \{ \tilde{X}_j(t) \}_{j = 1}^{N(n)} )  -  \frac{|S(\kappa_t)| } {1 - p} \right| \leq \frac{c(k)} {n},  \ n \geq 1.
\ee
\end{prop}

\bigskip
\noindent {\em Proof.}  We prove \eqref{rate} as \eqref{rateA} follows from identical methods.
We write $ \{X_j(t) \}_{j = 1}^{N(n)}$ as a Poisson point process $\P_{n \kappa_t}$ having intensity $n \kappa_t$ on $S(\kappa_t)$.  By the Mecke formula we have
\be \label{Mecke}
\E W_n^{(k)}(X_1(t),  \{X_j(t) \}_{j = 1}^{N(n)} ) = \int_{S(\kappa_t)} \E W_n^{(k)}(x,  \P_{n \kappa_t} ) \kappa_t(x) dx.
\ee
We let $\P_\tau$ be a Poisson point process on $\R$ of intensity $\tau$.
We have  $1 = \frac{2}{k} \E L^{(k)} (\0, \P_{1}  ) =  \frac{2}{k} \E L^{(k)} (x, \P_{1}  ) 
= \frac{2n }{k} \E L^{(k)} (x, \P_{n  \kappa_t(x) }  )  \kappa_t(x)$
 since  $n \tau \P_{n \tau}  \eqd \P_1$ for all $n \geq 1$ and all $\tau \in (0, \infty)$.
 Thus
 $$
  \E [ W_n^{(k)} (x, \P_{n \ka_t(x)})] \ka_t(x)  = 1.
 $$
 
\bigskip
Thus, integrating over all $x \in S(\kappa_t)$ gives
\be \label{newtransinv}  |S(\kappa_t)| =
 \int_{S(\kappa_t)   }  \E [ W_n^{(k)} (x, \P_{n \ka_t(x)})] \ka_t(x) dx.
\ee
Combining \eqref{Mecke} and \eqref{newtransinv} we get
\begin{align*}
& \left|  \E W_n^{(k)}(X_1(t),  \{X_j(t) \}_{j = 1}^{N(n)} )  - |S(\kappa_t)| \right| \\
& \leq
  \int_{S(\kappa_t) } \left| \E W_n^{(k)}(x,  \P_{n \kappa_t} ) -    \E W_n^{(k)} (x, \P_{n \ka_t(x)}) \right| \ka_t(x)  dx.
\end{align*}
Coupling arguments similar to those in the last section  of \cite{SY} show that there is a constant $c(k)$ such that
for all $x \in S(\kappa_t)$
$$
\left| \E W_n^{(k)}(x,  \P_{n \kappa_t} ) -    \E W_n^{(k)} (x, \P_{n \ka_t(x)}) \right|
\leq c [n^{-1} + \exp(-c n {\bf d} (x, \partial ( S(\kappa_t)  ) ) )],
$$
where $ {\bf d} (x, \partial ( S(\kappa_t)))$ stands for the distance between $x$ and the boundary of $S(\kappa_t)$. Combining the last two displays gives
$$
 \left|  \E W_n^{(k)}(X_1(t),  \{X_j(t) \}_{j = 1}^{N(n)} )  - 1 \right|
 \leq c \int_{S(\kappa_t)} [n^{-1} +\exp(-c n {\bf d} (x, \partial ( S(\kappa_t) ) ))] \ka_t(x)  dx.
 $$
Making a change of variable gives the desired result.  \qed

\bigskip

\bigskip

Now we are ready to give the proof of  Theorem \ref{CLT}. 

 \bigskip

\noindent {\em Proof.}  We first establish the asymptotic normality assertions. 
To establish \eqref{markedCLTfixt}  we first show  as $n \to \infty$
\be \label{CLTfixtA}
\frac{  \sum_{i \in I_0(t)} \Big( W_n^{(k)}\Big(\tilde X_i(t),  \{\tilde X_j(t) \}_{j = 1}^{N(n)}\Big) - \E W_n^{(k)}\Big(\tilde X_i(t),  \{\tilde X_j(t) \}_{j = 1}^{N(n)}\Big) \Big)} { \sqrt{n(1-p)}} \tod  N \bigg(0, \nu^2(t,k) \bigg).
\ee
The limit  \eqref{CLTfixtA}  is a consequence of general limit theory for sums of exponentially stabilizing functionals on marked Poisson point sets, as given in e.g.
\cite{BY05} and \cite{PenEJP}.   It suffices to note that the localization distance $L^{(k)}$ is an  exponentially stabilizing score function.  This is because its value at a point $x$ is determined by the spatial locations of the $k$ nearest neighbors to $x$ and because the density $\ka_t$ is bounded away from zero.  Such functionals are known to be stabilizing, see e.g. \cite{LSY} and \cite{PY4}.
To deduce \eqref{markedCLTfixt} from  \eqref{CLTfixtA} we apply the rate result \eqref{rateA}. 

\bigskip
\sloppy We deduce the rate results \eqref{markedCLTrate} and  \eqref{markedCLTrate2}
from a general result on rates of normal convergence for exponentially stabilizing functionals of  marked point processes; see Theorem 2.3(a) of \cite{LSY}. In particular we make use of the growth bounds  $\Var [ \sum_{i = 1}^{N(n)}  W_n^{(k)}(\tilde X_i(t),  \{\tilde X_j(t) \}_{j = 1}^{N(n)})] = \Theta(n)$, the validity of which is discussed in Remark 2, following Theorem 3.1 in \cite{LSY}.  The lower bound  $\Var [ \sum_{i = 1}^{N(n)}  W_n^{(k)}(\tilde X_i(t),  \{\tilde X_j(t) \}_{j = 1}^{N(n)})] = \Omega(n)$ insures that 
 \be \label{markedCLTrateAA}
d_K \left( \frac{  \sum_{i \in I_0(t)} \Big( W_n^{(k)}\Big(\tilde X_i(t),  \{\tilde X_j(t) \}_{j = 1}^{N(n)} \Big) - \E W_n^{(k)}\Big(\tilde X_i(t),  \{\tilde X_j(t) \}_{j = 1}^{N(n)}\Big) \Big)}
 { \sqrt{ \Var  \sum_{i \in I_0(t)}  W_n^{(k)}\Big(\tilde X_i(t),  \{\tilde X_j(t) \}_{j = 1}^{N(n)} \Big)   }},  N(0,1) \right) \leq \frac{ C_1(k) } {\sqrt {n}}.
\ee 
  The upper bound  $\Var [ \sum_{i = 1}^{N(n)}  W_n^{(k)}(\tilde X_i(t),  \{\tilde X_j(t) \}_{j = 1}^{N(n)})] = O(n)$, along with \eqref{rateA}, insure that replacing 
 $ \E W_n^{(k)}\Big(\tilde X_i(t),  \{\tilde X_j(t) \}_{j = 1}^{N(n)}\Big)$ by  $(1 - p)^{-1}$
 gives an error which is at most $\frac{ C(k) } {\sqrt {n}}$  where $C(k)$ is a constant which depends on $k$. These two remarks give 
 \eqref{markedCLTrate}.  The rate  \eqref{markedCLTrate2} is proved similarly.

The asymptotics \eqref{markedVart} may be deduced from general variance asympotics for sums of exponentially stabilizing score functions on marked Poisson input.  We refer the reader to \cite{BY05} and also \cite{PenEJP}. 
The limit   \eqref{markedratio} follows since
$$
\lim_{n \to \infty} \frac{ \Var  \sum_{i \in I_1(t)}  W_n^{(k)}\Big(\tilde X_i(t),  \{\tilde X_j(t) \}_{j = 1}^{N(n)} \Big) } { np  } =  \frac{4 \tilde \sigma_t^2}{(1-p)^2k^2}.
$$
This completes the proof of Theorem \ref{CLT}. 
 \qed

 \section{Acknowledgements}
The research of J. Yukich is supported in part by a Simons collaboration grant.  He is also grateful for generous support from the Department of Statistics at Universidad Carlos III de Madrid, where most of this work was completed.

\end{document}